\documentclass[twocolumn,longbibliography,showpacs,superscriptaddress,twoside,pra]{revtex4-1}
\usepackage{amsmath,amssymb,url,graphicx,epstopdf,bm}
\usepackage{epsfig,hyperref,amsmath,amssymb,amsfonts, latexsym, dsfont, wasysym,multirow,mathrsfs,xcolor,color}
\usepackage[caption=false]{subfig}
\usepackage{braket}

\begin{document}
\title{Designing High-Fidelity Single-Shot Three-Qubit Gates: A Machine Learning Approach}
\author{Ehsan Zahedinejad}
\affiliation{Institute for Quantum Science and Technology, University of Calgary, Alberta, Canada T2N 1N4}
\author{Joydip Ghosh}\email{jghosh3@wisc.edu}
\affiliation{Institute for Quantum Science and Technology, University of Calgary, Alberta, Canada T2N 1N4}
\affiliation{Department of Physics, University of Wisconsin-Madison, Madison, Wisconsin 53706, USA}
\author{Barry C. Sanders}\email{sandersb@ucalgary.ca}
\affiliation{Institute for Quantum Science and Technology, University of Calgary, Alberta, Canada T2N 1N4}
\affiliation{Program in Quantum Information Science, Canadian Institute for Advanced Research, Toronto, Ontario M5G 1Z8, Canada}
\affiliation{Hefei National Laboratory for Physical Sciences at the Microscale and Department of Modern Physics, University of Science and Technology of China, Anhui 230026, China}
\affiliation{Shanghai Branch, CAS Center for Excellence and Synergetic Innovation Center in Quantum Information and Quantum Physics, University of Science and Technology of China, Shanghai 201315, China}
\affiliation{Institute for Quantum Information and Matter, California Institute of Technology, Pasadena, California 91125, USA}
\begin{abstract}
Three-qubit quantum gates are key ingredients for quantum error correction and quantum information processing.
We generate quantum-control procedures to design three types of three-qubit gates, namely
Toffoli, Controlled-Not-Not and Fredkin gates.
The design procedures are applicable to a system comprising three nearest-neighbor-coupled superconducting artificial atoms.
For each three-qubit gate, the numerical simulation of the proposed scheme achieves 99.9\% fidelity, which is an accepted threshold fidelity for fault-tolerant quantum computing.
We test our procedure in the presence of decoherence-induced noise as well as show its robustness against random external noise generated by the control electronics. The three-qubit gates are designed via the machine learning algorithm
called Subspace-Selective Self-Adaptive Differential Evolution (SuSSADE).
\end{abstract}

\date{\today}
\maketitle
\section{Introduction}
Quantum computing requires a universal set of low-error quantum gates to enable fault-tolerant
quantum computing~\cite{Got98},
and characterization is typically performed using average fidelity from which gate error rate can be inferred~\cite{SWS16}.
A universal set of single-~\cite{MGR+09} and two-qubit~\cite{BKM+14} gates can be employed to decompose~\cite{MMJ+04} multi-qubit gates into a series of single and two-qubit gates. In practice, this decomposition-based approach
is undesirable because it leads to quantum circuits~\cite{CPM+98,RDN+12} with long operation time. We employ the recently proposed machine-learning technique, 
which we named Subspace-Selective Self-Adaptive Differential Evolution (SuSSADE)~\cite{ZGS15},
to generate \emph{procedures} (i.e., set of instructions that determine the control parameters, and hence the effectiveness of the control scheme) for designing single-shot high-fidelity three-qubit gates without any need to resort to a decomposition.
We test our procedure~\cite{HS10} in the presence of decoherence induced noise and demonstrates its robustness under the effect of random control noise. 
The three-qubit gates that we consider here are Toffoli~\cite{MKH+09,FSB+12,SFW+12} (which has already been discussed in Ref.~\cite{ZGS15} and we give a review here for completeness), Fredkin~\cite{Mil89,JHW+86} and Controlled-NOT-NOT or CXX~\cite{GFG12}, which are typical three-qubit gates employed for quantum information processing.

The three-qubit gates that we consider in this study are key ingredients for quantum algorithms and error correction. A quantum Toffoli gate is necessary for (non-topological) quantum error correction~\cite{CPM+98,RDN+12} and a key component for reversible computing~\cite{Per85}. Toffoli gate together with single-qubit Hadamard gate comprise a universal set of quantum gates~\cite{Aa03,Shi03}. The Fredkin gate enables reversible computing~\cite{Per85}, and also forms a universal set along with Hadamard gate~\cite{SW95}. The CXX gate appears in the syndrome operator measurement circuit~\cite{Wei15} for quantum error correction algorithms, such as the Steane code~\cite{Ste96} and surface code~\cite{GFG12}. Although the Fredkin and CXX gates can be decomposed into three and two CX (i.e., CNOT) gates respectively~\cite{BKM+14}, we avoid the decomposition-based approach and generate a procedure to design these three-qubit gates that achieve the same gate action over a shorter timescale.

Thus far Toffoli and Fredkin gates are achieved by decomposition into single- and two-qubit gates~\cite{CW95,SD96,SM09} in various physical systems, and yet none of these efforts have achieved the threshold fidelity~\cite{BKM+14}.
Recently we have proposed a quantum control scheme (called SuSSADE) for designing a single-shot high-fidelity ($>$ 99.9\%) Toffoli gate for a system comprising three nearest-neighbor-coupled superconducting artificial atoms~\cite{ZGS15}. In this work we show that this machine learning
technique enables the design of other three-qubit gates as well for the same physical model. For all the three-qubit gates considered in this work we show that the gates operate as fast as a two-qubit entangling Controlled-Z (CZ) gate under the same experimental constraints.

Recent progress in superconducting artificial atoms~\cite{GPS+07,DWM04,CW08} has made them appealing for quantum information processing, especially for gate-based quantum computing~\cite{Rig11}. An avoided-crossing-based CZ gate~\cite{GGZ+13} has
already been achieved for a system comprising coupled superconducting artificial atoms.'
The idea of avoided-crossing-based gates is to vary the energy level of artificial atoms such that energy levels approach each other but avoid degeneracies.
These avoided crossings mix population and dynamical phases at the corresponding levels such that the final evolution of the system gives the required phase and population for the target gates. 

Our strategy to design the three-qubit gates is also based on the avoided level-crossings. We employ an open-loop quantum control approach to generate optimal external pulses for the frequencies of the superconducting artificial atoms. We employ machine learning as a quantum control tool to generate successful procedures and show that our procedure can also be implemented in an open quantum system where external noise is also acting on the system.

Other approaches exist to design quantum gates for a network of superconducting transmon systems, where one can couple each transmon with a microwave generator~\cite{PhysRevB.81.134507} or couple the transmons via tunable couplers~\cite{PhysRevLett.113.220502} and control these external circuit elements to evolve the system toward a specific unitary operation. These approaches require more resources (additional circuit elements) compared to our approach~\cite{ZGS15}, where we only control the transmon frequency via a quantum control scheme~\cite{SB03,DP10} and evolve the system's dynamics toward the target gate.

Machine learning~\cite{SB98} is concerned with the construction of algorithms that can learn from data and make predictions on data.
Typical machine learning algorithms tend to be greedy~\cite{MSG+11,DCM11},
as they need less resource and computational time to complete the learning task as well as converge faster (in comparison to non-greedy approaches~\cite{BYW+97,ZSS14}). However, we have observed that greedy machine-learning techniques failed to generate a successful procedure for designing high-fidelity three-qubit gates, which motivates us to employ the non-greedy machine-learning technique. Our learning algorithm is based on an enhanced version of Differential Evolution (DE) algorithm~\cite{SP97},
hence the name we assigned:
Subspace-Selective Self-adaptive DE or SuSSADE~\cite{ZGS15}.

The rest of the paper is organized as follows. In Section~\ref{sec:model} we explain the physical model that we use to design the three-qubit gates. In Section~\ref{sec:AvoidedCross} we discuss the avoided-crossing-based gates for two- and three-qubit gates. In particular, we review the current theoretical framework for designing avoided-crossing-based CZ gate and also discuss that why taking the same theoretical approach is challenging for avoided-crossing-based three-qubit gates. In Section~\ref{sec:QC} we discuss our quantum control scheme and show that how we translate the problem of designing a three-qubit gate into a learning algorithm. In Section~\ref{sec:noiseModel} we discuss the noise model. In Section~\ref{sec:threeQubitLogicalGates} we discuss each individual three-qubit gate and giving their effect on a quantum state. Section~\ref{sec:result} presents the results. The significance of the results is outlined in Section~\ref{sec:discussion} and we conclude our work in Section~\ref{sec:conclusion}.

\section{Physical Model}
\label{sec:model}
We consider a system comprising three nearest-neighbor-coupled superconducting artificial atoms~\cite{BKM+14} with parameters appropriate for the transmon system~\cite{KYG+07,HKD+09}. Each transmon is capacitively coupled to its nearest neighbor, where the location of each transmon is labeled by $k=1, 2, 3$. The frequency $\varepsilon_k(t)$, in the rotating-frame, can be tuned via superconducting control electronics.
The anharmonicities of the second and third energy levels are represented by~$\eta$ and $\eta'$.

We approximate $\eta'=3\eta$,
which is valid for the cubic approximation of the potential for the transmon system~\cite{GGZ+13}.
The transmons are coupled capacitively,
which yields an~$XY$ interaction between adjacent transmons (in the rotating frame)
with a coupling strength~$g$.
The Hamiltonian for three capacitively coupled transmons is thus~\cite{GGZ+13}
\begin{align}
\label{eq:qutritcontrol}
	\frac{\hat{H}(t)}{h}
		=&\sum_{k=1}^{3}
			\begin{pmatrix} 0&0& 0&0\\0&\varepsilon_k(t)&0&0\\0&0&2\varepsilon_k(t)-\eta&0\\0&0&0&3\varepsilon_k(t)-\eta'\end{pmatrix}_{k}
						\nonumber\\&
		+\frac{g}{2}\sum_{k=1}^{2}(\hat{X}_k\hat{X}_{k+1}+\hat{Y}_k\hat{Y}_{k+1}),
\end{align}
where
\begin{equation*}
		\frac{\hat{Y}_k}{i}
			=\begin{pmatrix} 0&-1& 0&0\\1&0&-\sqrt{2}&0\\0&\sqrt{2}&0&-\sqrt{3}\\0&0&\sqrt{3}&0\end{pmatrix}_{k},
		\hat{X}_k=\begin{pmatrix} 0&1& 0&0\\1&0&\sqrt{2}&0\\0&\sqrt{2}&0&\sqrt{3}\\0&0&\sqrt{3}&0\end{pmatrix}_{k}
\end{equation*}
are the generalized Pauli operators~\cite{GGZ+13}.

The experimental constrains for the transmons require specific values for each physical parameter in Eq.~(\ref{eq:qutritcontrol}). The transmon frequencies $\varepsilon_k(t)$ are varied between $2.5$ and $-2.5$ GHz. We consider
\begin{equation}
	\eta=200~\text{MHz},
	g=30~\text{MHz}.
\end{equation}
Although the physical system considered for this work consists of superconducting circuits, our quantum control scheme is, however, not limited to a specific system. 

The Hamiltonian~(\ref{eq:qutritcontrol}) generating the three-qubit gates
acts on a $4^3$-dimensional Hilbert space
$\mathscr{H}_4^{\otimes 3}$.
Under the rotating-wave approximation, this Hamiltonian
is a block-diagonal matrix with each block corresponding to a fixed number of excitations.
This block diagonalization property permits us to reduce the $4^3$-dimensional Hamiltonian to a subspace where at most $3$ excitations are present,
which is the relevant subspace for three-qubit gates.

We define a projection operator $\mathscr{O}_m$ that truncates the Hamiltonian~(\ref{eq:qutritcontrol}) up to the $m^\text{th}$ excitation subspace, and for $m=3$ (that allows at most 3 excitations,
which is all what we need for three-qubit gates) the projected Hamiltonian is given by,
\begin{equation}
\label{eq:Hp}
\hat{H}_\text{p}(t)=\mathscr{O}_3\hat{H}(t)\mathscr{O}_3^{\dagger}.
\end{equation}
We observe that the unitary evolution of the system is unaffected by this truncation, which is what we expect from a block-diagonal Hamiltonian.

We evolve ${\hat{H}_\text{p}(t)}$ such that the resultant unitary operator is
\begin{equation}
\label{eq:U}
	U(\Theta)
		=\mathcal{\hat{T}}\exp\left\{-\text{i}\int_0^\Theta\hat{H}_\text{p}(\tau)\text{d}\tau\right\}
\end{equation}
with~$\mathcal{\hat{T}}$ the time-ordering operator~\cite{DGT86}. 
Note that $U(\Theta)$ is a $20\times20$ unitary operator whereas the three-qubit gates reside in $2^3$-dimensional computational subspace. We therefore, define another projection operator $\mathscr{P}$, which projects $U(\Theta)$ into the computational subspace of the three-transmon system
\begin{equation}
\label{eq:pUp}
	U_\text{cb}(\Theta)=\mathscr{P}U(\Theta)\mathscr{P}^{\dagger},
\end{equation}
where $U_\text{cb}$ is the projected unitary operator.

Our goal is to achieve these specific three-qubit unitary operation (Toffoli, Fredkin or CXX)  over the duration $\Theta$, such that the distance between $U_\text{cb}(\Theta)$ and the target three-qubit gate is minimal.
We evolve the system Hamiltonian~(\ref{eq:U}) such that the final time evolution operator approaches to the target three-qubit gate modulo some phases that can be compensated by local $z$-rotations on each transmon.
This phase compensation~\cite{GKM12,GGZ+13} is performed via the excursions of transmon frequencies and is trivial for superconducting circuits.

To steer the system dynamics towards a specific entangling gate operation, we define the equivalence class of a given three-qubit gate $U_{\text{target}}$ under local $z$-rotations as~\cite{GGZ+13}
\begin{equation}
	U_{\text{target}}
		\equiv U'_{\text{target}}
		=U_{\text{post}} U_{\text{target}} U_{\text{pre}},
\end{equation}
where
\begin{equation}
\label{eq:eqv}
U_{\text{pre,post}}(\beta_1,\beta_2, \beta_3) \equiv R_z(\beta{_1})\otimes R_z(\beta{_2})\otimes R_z(\beta{_3}).
\end{equation}
$R_{z}$ in~(\ref{eq:eqv}) denotes a unitary single-qubit rotation about the z-axis. Equation~(\ref{eq:eqv}) can be explicitly expressed in terms of $\{\beta_j\}$,
which is the set of local phases acquired by the $j^\text{th}$ transmon:
\begin{align}
\label{eq:smatrix}
	U_{\text{pre,post}}
		=&\text{diag}\left(1,\text{e}^{-\text{i}\beta_3},\text{e}^{-\text{i}\beta_2},
		\text{e}^{-\text{i}\left(\beta_2+\beta_3\right)},\text{e}^{-\text{i}\beta_1},\right.\nonumber\\
	&\left.\text{e}^{-\text{i}(\beta_1+\beta_3)},\text{e}^{-\text{i}(\beta_1+\beta_2)},\text{e}^{-\text{i}(\beta_1+\beta_2+\beta_3)}\right).
\end{align}
We use $U_{\text{pre,post}}$, which are diagonal $8\times8$ matrices operating on $2^3$ dimensional computational subspace of the three qubits, to perform phase compensation in the numerical simulation of each three-qubit gate.

\section{Avoided-Crossing-Based gates}
\label{sec:AvoidedCross}
In this section, we first discuss the avoided-crossing-based technique in designing the two-qubit entangling CZ gate.
Such a scheme was first proposed for a system of two coupled phase qubits~\cite{SJD+03},
which was later adapted for a system of resonator-coupled superconducting qubits~\cite{GGZ+13}. We describe the avoided-crossing-based CZ gate for the physical model of two capacitively coupled frequency-tunable transmons, with~$\eta$ and~$g$ ($\ll \eta$) being the anharmonicity and the coupling strength, respectively. This discussion is necessary to clarify why finding a theoretical solution for three-qubit gates is challenging, which is in fact the motivation for our quantum control approach.

\subsection{CZ gate based on avoided level-crossing}
\label{subsec:CZ}
CZ gate is a two-qubit entangling gate. It exerts a Pauli $z$-rotation on the second (target) qubit if and only if the first (control) qubit is~$\ket{1}$. The CZ gate acts on the basis states according to
\begin{align}
\label{eq:CZmap}
\ket{00}\mapsto&\ket{00},\;
\ket{01}\mapsto\ket{01},\nonumber\\
\ket{10}\mapsto&\ket{10},\;
\ket{11}\mapsto-\ket{11},
\end{align}
which leaves three two-qubit basis states intact and imposes a sign change on one basis state.

\begin{figure}
	\includegraphics[width=\columnwidth]{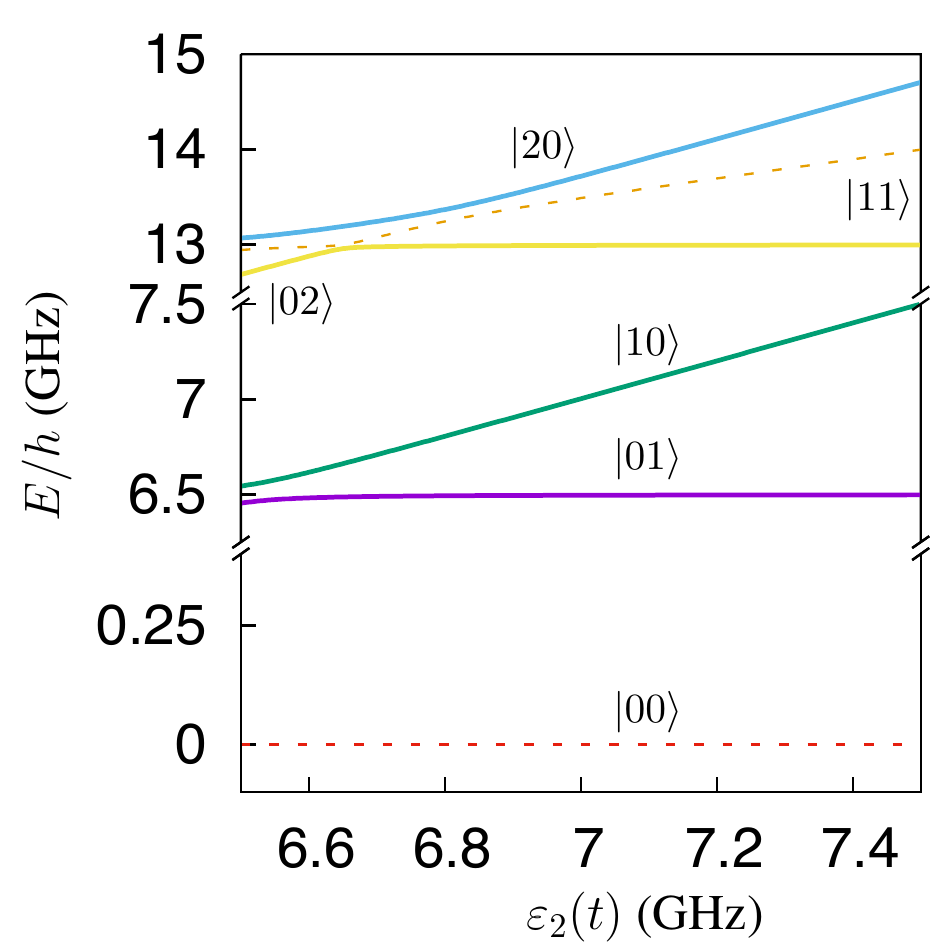}
\caption{
	(color online)
	The energy ($E$) spectrum of the system Hamiltonian of two capacitively coupled transmons.
	The frequency of the first transmon is fixed at $6.5$ GHz. The frequency~$\varepsilon_2(t)$
	of the second transmon varies from $7.5$ to $6.5$ GHz}
\label{fig:Avoided_CZ}
\end{figure}

The energy levels of two capacitively coupled transmons are shown in Fig.~\ref{fig:Avoided_CZ}, where we fix the frequency of the first transmon at $\varepsilon_1=6.5$ GHz and allow the frequency of the second transmon $\varepsilon_2(t)$ to vary from $\omega_\text{off}=7.5$ GHz (detuned frequency) to
\begin{equation}
	\omega_\text{on}=\varepsilon_1+\eta=6.7~\text{GHz.}
\end{equation}
We vary the frequency of the second transmon using a time-dependent error function, with a switching time of $t_\text{ramp}$ and the gate operation time of $t_\text{gate}$.
This time-dependent frequency is~\cite{GGZ+13},
\begin{align}
\varepsilon_2(t)
=&\omega_\text{off}+\frac{\omega_\text{on}-\omega_\text{off}}{2}\left[\operatorname{erf}{\left(\frac{4t-2t_\text{ramp}}{t_\text{ramp}}\right)}\right.\nonumber\\
&-\left.\operatorname{erf}{\left(\frac{4t-4t_\text{gate}+2t_\text{ramp}}{t_\text{ramp}}\right)}\right].
\end{align}

The avoided-crossing-based CZ gate works as follows: Initially, we detune the transmon frequencies from each other by setting the frequency of the second transmon equal to $\varepsilon_2(t=0)=\omega_\text{off}$. This makes all the eigenstates of the system non-degenerate. Then we tune the second transmon to $\omega_\text{on}$ for a time
\begin{equation}
	t_\text{on}=t_\text{gate}-2t_\text{ramp},
\end{equation}
and finally detune the second transmon again to the frequency $\omega_\text{off}$. During the time $t_\text{on}$, the computational basis state $\ket{11}$ mixes with the two auxiliary levels $\ket{02}$ and $\ket{20}$, whereas all the other eigenstates in the computational basis ($\ket{00}$,~$\ket{01}$,~$\ket{10}$) are detuned from each other. The parameters of the control pulse are determined such that the mixing among $\ket{11}$,~$\ket{02}$ and~$\ket{20}$ states over the time interval $t_\text{on}$ ensures the phase factors required for the CZ gate as shown in Eq.~(\ref{eq:CZmap}).

Depending on the timescale $t_\text{on}$,
two distinct regimes exist in which a CZ gate can operate: 
the sudden-approximation regime and the adiabatic regime.
In the sudden-approximation regime we vary the second qubit frequency fast enough, such that the switching time can be made sudden with respect to~$g$ (but still adiabatic with respect to~$\eta$). For the sudden-approximation regime, the switching time has an inverse relation with the coupling factor as
\begin{equation}
t_\text{on}=\frac{\pi}{\sqrt{2}g}.
\label{eq:CZtime}
\end{equation}
Under the sudden approximation,
two parameters of the pulse
$\omega_\text{on}$ and $t_\text{on}$
must be optimized to obtain a high-fidelity CZ gate.

On the other hand, in the adiabatic regime, the switching is adiabatic with respect to~$g$ (and therefore, with respect to~$\eta$ as well). In this regime, stronger coupling is required between transmons to make the gate operate as fast as the gates in the sudden approximation regime. This increase of~$g$ leads to residual errors in a multiqubit device, whereas the advantage of operating the transmon in the adiabatic regime is that it suffices to optimize only one parameter. This is because the adiabatic regime ensures that the population of each energy level is preserved. A CZ gate in the adiabatic regime is demonstrated in Ref.~\cite{DCG+09}.

\subsection{Avoided-crossing-based approach for three-qubit gates}
Either in sudden or in adiabatic regime, the idea of engineering a pulse for the avoided-crossing-based gate in a superconducting system remains the same: Designing a control pulse for the qubit frequency, such that the $\ket{11}$ state mixes with the other states in the second excitation subspace, while the states in zero- and single-excitation subspaces remain detuned from each other. However, for practical implementations,
both sudden and adiabatic regimes are unsuitable for obtaining the threshold fidelity required for fault-tolerance.
Instead advanced machine-learning-based techniques can  be employed to engineer optimal pulses, which is the  motivation for our work. 

One idea for designing three-qubit gates is to couple three transmons via a superconducting cavity, usually referred to
as the circuit-quantum-electrodynamics (cQED) architecture~\cite{WSB+04}, and tune the transmon frequencies in the dispersive regime such that the time-evolution operator gives rise to the target three-qubit at the end of operation. Such an approach has already been used to demonstrate a Toffoli gate~\cite{RDN+12}. We, however, do not consider the cQED hardware in our work, as the architecture can only contain a few transmons inside a superconducting cavity, and therefore, is not scalable.

\begin{figure}
	\includegraphics[width=\columnwidth]{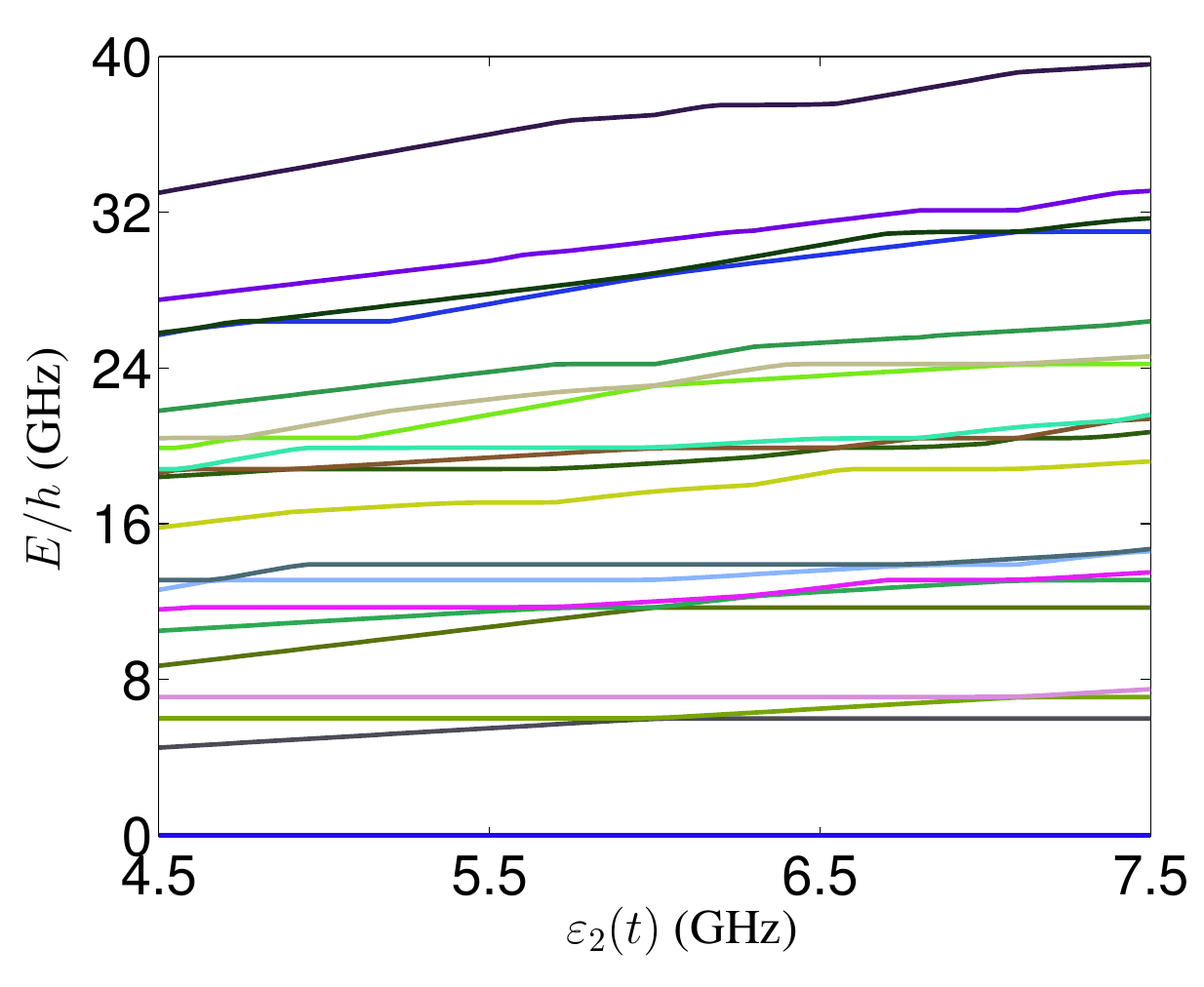}
\caption{
	(color online)
	The energy ($E$) spectrum of three nearest-neighbor-coupled transmons. The first and third transmon frequencies are fixed at $4.8$ and $6.8$ GHz respectively. The frequency of the second transmon varies from $4.5$ to $7.5$ GHz}
\label{fig:Avoided_C_CZ}
\end{figure}

Instead we consider a one-dimensional chain of three transmons with nearest-neighbor coupling. To see if such an approach is suitable for avoided-crossing-based three-qubit gates, we first plot the energy spectrum of such a three-transmon system in Fig.~\ref{fig:Avoided_C_CZ},
where we fix the frequencies of the first and third transmons to 4.8 and 6.8~GHz, respectively, and vary the frequency of the second one from 4.5  to 7.5~GHz.
We also set
\begin{equation}
	g=30~\text{MHz},\;
	\eta=200~\text{MHz}.
\end{equation}

In contrast to the two-qubit case, the energy spectrum of a three-transmon system is crowded with many level-crossings (or anticrossings in the presence of interaction). For this system, therefore, finding optimal pulses theoretically for the transmon frequencies, such that the $\ket{111}$ state mixes strongly with the other states in the third excitation subspace, whereas all the other states are detuned from each other is a challenging task, which is why we employ the quantum control scheme and in particular the machine-learning technique to devise a procedure for designing such three-qubit gates~\cite{ZGS15}.

\section{quantum control}
\label{sec:QC}
In this section we first give the application of quantum control and elaborate how a gate-design problem can be transformed into a quantum-control problem. We introduce supervised machine learning as a tool to generate procedures for designing quantum gates.
We explain the map from quantum-control problem to the learning problem in the second part of this section, and then show that generating a successful procedure via the learning algorithm becomes a feasibility problem for which optimization algorithms can be employed. We discuss some existing optimization algorithms that we employed to find feasible procedures, discuss how these algorithms remain inadequate to yield the required fidelity, and then introduce the enhanced version of differential evolution algorithm.

\subsection{quantum-gate design as control problem}
In the context of optimal control theory, the main task of quantum control is to investigate how to steer quantum dynamics towards a specific quantum state or operation~\cite{DP10}. The emergence of new quantum technologies have realized new applications for quantum control in various fields, such as femtosecond lasers~\cite{ABB+98,MS98}, nuclear magnetic resonance and other resonators~\cite{NMR1,KRK+05,RSK05,MVT98,HJH+03}, laser-driven molecular reactions~\cite{BS92,TR85}, and quantum gate synthesis for quantum computing~\cite{SSK+05}. In particular, we employ quantum control scheme to design fast and high-fidelity three-qubit gates.

As we explained earlier, a chain of three capacitively coupled transmons constitute the physical model for the three-qubit gates. This physical system evolves according to Eq.~(\ref{eq:U}) and our goal is to steer the evolution toward the target three-qubit gate operation. In order to turn the problem into a quantum-control problem, we need to clarify what the control parameters are. In our control scheme, control parameters are the qubit frequencies which can be tuned via external pulses. The task of finding the optimal shape for the external pulses can be performed via quantum control schemes. Machine learning can be employed as a quantum control tool to perform this task.

\subsection{Supervised machine learning: a quantum control tool}
The task of machine learning~\cite{Bis06} is to develop algorithms which can learn from system behaviour and predict the future behaviour of the system based on their past evolution. Machine learning algorithms have already been applied to various problems in quantum information science, such as phase estimation~\cite{HS10}, asymptotic state estimation~\cite{MW10}, and discriminating quantum measurement trajectories and improving readout~\cite{MGM+15}. One can classify the machine learning algorithms in three distinct categories namely, supervised learnings, unsupervised learnings and reinforcement learnings~\cite{Bis06}. Our focus is on supervised machine learning algorithm as a quantum control tool.

A supervised learning task~\cite{Kot07} is to infer a function (hypothesis) from the labeled data (training set). The training data comprises an input vector accompanied by its corresponding output vector. 
A supervised learning algorithm trains the hypothesis on the training data to construct an inferred hypothesis. This inferred hypothesis can be further used to label novel data. Examples of supervised learning problems are regression or classification problems~\cite{CN06,BPB15}. The idea of supervised learning can be generalized to develop quantum control schemes that deliver successful procedures for quantum-gate design. In what follows, we describe the procedure for turning a gate-design problem into a supervised machine learning problem.

A quantum logic gate is a map between an input and an output state. One can always represent the action of any quantum logical gate on the basis elements in terms of a truth table. There is a one-to-one correspondence between the input and output elements in this truth table, as the quantum logic gates are themselves reversible. In the context of supervised learning problem, we consider the truth table as the training set. Loosely speaking, we train our hypothesis on the truth table data as the training set.

Having clarified that the truth table represents the training set, we now discuss what the hypothesis is.  In the context of quantum-gate design, the hypothesis is the external pulses. Therefore, we train the parameters of the external pulses on the truth table data to shape the external pulses such that the system evolution approximates the target gate. If the hypothesis is learnt successfully, it generates a procedure which determines the shape and strength of the external pulses. Thus far, we have explained the training set and hypothesis in the context of quantum-gate design. However, we still need to know how to measure the success of our learning procedure as well as the explicit form of the hypothesis (external field), which we elaborate in the next two subsections.

\subsection{Control pulses as learning parameters}
In an avoided-crossing-based gate, the energy levels of the artificial atoms are varied using the external pulses to steer the dynamics of the system toward the desired operation. Our learning algorithm uses these external pulses as the learning parameters. Although the quantum control approach can generate any type of pulses for $\varepsilon_k(t)$, we only consider two types of pulses: piecewise-constant and piecewise-error-function.

In the case of piecewise-constant function we discretize each transmon frequency, $\varepsilon_k(t)$, with $k\in$\{1,2,3\} and express it as sum of $N$ orthogonal constant functions over the interval [0, $\Theta$]. Each $\varepsilon_k$ then can be shown as
\begin{equation}
\label{eq:piecewise}
	\varepsilon_k
		:=
		\begin{pmatrix}
			\varepsilon_{k_1} \\ \varepsilon_{k_2}\\ \vdots \\ \varepsilon_{k_N}
		\end{pmatrix}
\end{equation}
with each $\varepsilon_{k_l}$ the magnitude of the $k^\text{th}$ pulse at the $l^\text{th}$ time step. 
The time bins are chosen to be equally spaced over the interval $\Theta$ and is therefore, given by
\begin{equation}
\Delta{t}=\frac{\Theta}{N-1}.
\end{equation}
The pulse generators for superconducting devices can generate the piecewise-constant functions. However, such rectangular-shaped pulses get distorted by the Gaussian filters that bridge the control circuitry with the transmons. More realistic pulse shapes that can take such distortions into account should therefore, be considered.

In order to consider the distortion on the rectangular pulses, we connect each control parameter in~(\ref{eq:piecewise}) by including following error-function coefficient~\cite{GGZ+13} in
\begin{align}
\varepsilon_{k}(t)=\frac{\varepsilon_{k_l}+\varepsilon_{k_{l+1}}}{2}+&\frac{\varepsilon_{k_{l+1}}-\varepsilon_{k_l}}{2}\nonumber\\&
\text{erf}\left[\frac{5}{\Delta{t}}\left(t-\frac{t_l+t_{l+1}}{2}\right)\right]
\label{eq:erfPulse}
\end{align}
for $t_l \leq t \leq t_{l+1}$, $t_l$ being the $l^\text{th}$ time step. The resultant smooth pulse expressed in~(\ref{eq:erfPulse}) accounts for the first-order distortion caused by Gaussian filters. In Sec.~\ref{subsec:Cpulsedist} we explain how one can implement an adaptive control loop
to suppress the higher-order noise.

In general, designing smooth pulses are computationally more expensive than computing for piecewise-constant functions.
We therefore, choose less expensive piecewise-constant functions to analyze the gate fidelity against various parameters but could incorporate other shapes if the extra computational cost is warranted.
We justify this choice by showing that the gate fidelity does not depend on the type of the pulse, but depends on the number of learning parameters.

\subsection{Confidence or fitness functional}
The standard method to measure the performance of a supervised learning algorithm is to define a confidence for the learnt hypothesis. The confidence is the ratio of the number of training data that are learnt successfully to the total number of training data.
If the learning task is to generate a procedure for designing a quantum gate, one can define this confidence,~$\mathcal{F}$, by the distance between target and approximated unitary operators:
\begin{equation}
\label{eq:F}
	\mathcal{F}=
		\|\mathscr{P}U(\Theta)\mathscr{P}^{\dagger}-U_\text{target}\|
\end{equation}
with~$\|\bullet\|$ the operator norm so $\mathcal{F}$ is
the trace distance~\cite{GVC12} between the target and actual evolution
operators projected to the computational subspace.

Our machine learning algorithm uses an explicit form of~(\ref{eq:F}), where
\begin{equation}
\label{eq:explicitF}
\mathcal{F}
		=\frac{1}{8} \left|\text{Tr}\left(U_\text{target}^{\dagger}\;U_\text{cb}(\Theta)\right)\right|.
\end{equation}
Note that $\mathcal{F}=1$ if $U_\text{cb}(\Theta)$ corresponds to the target three-qubit gate $U_\text{target}$ (Toffoli, Fredkin or CXX) and $0\leq\mathcal{F}<1$ otherwise.

In the context of quantum-gate design, the confidence functional $\mathcal{F}$ is called the intrinsic fidelity which refers to the fidelity between the unitary evolution of the closed quantum system
of~(\ref{eq:qutritcontrol}) and the target operation when the decoherence noise is ignored. We follow the standard practice of gate design
by first considering a closed quantum system and generating a successful procedure for the learning task, and then evaluate the performance of generated procedure in the presence of noise. We call the confidence of our learnt hypothesis in the presence of noise as average state fidelity $\bar{\mathcal{F}}$~\cite{ZGS15}, which is
\begin{equation}
\label{eq:Fbar}
\bar{\mathcal F}:=\frac{1}{8}\sum_k\sqrt{\left|\bra{\psi_k}\rho^\text{final}_k\ket{\psi_k}\right|},
\end{equation}
where $\rho^\text{final}_k$ is the final density matrix of the system and $\ket{\psi_k}$ is the $k^{\text{th}}$ basis state in the computational subspace.

\subsection{Machine Learning and optimization algorithms}
\label{sec:SuSSADE}
For a gate-design problem, one can turn the problem of finding a successful procedure into a feasibility problem by setting the fidelity between the obtained unitary operation and the target gate to a fixed value that is acceptable by fault-tolerant quantum computing. All procedures that result in the error within the threshold value are called feasible (successful) procedures. Various optimization algorithms can be employed to tackle this feasibility problem.

Greedy algorithms~\cite{SdF11,MSG+11,DCM11} are at the heart of the machine-learning techniques. Given a quantum control system with no physical constraints, finding a feasible procedure is a trivial task for greedy algorithms~\cite{RHR04}. However, we have already shown~\cite{ZSS14} that reducing either gate operation time~$\Theta$ or number of control parameters raises the difficult of the quantum-control problem. One might need to try different optimization algorithms to generate procedures for designing quantum gates under constraints of operation time and experimental resource. 

We have examined the existing quantum control schemes including the quasi-newton~\cite{Bro70,Fle70,Fle13,Gol70,Sha70} and Nelder-Mead~\cite{ON75} algorithms (greedy) as well as DE~\cite{SP97} and Particle Swarm Optimization~\cite{CK02} algorithms (non-greedy). We observed that all these schemes fail to generate a fidelity better than 99.5\% (summarized in Table \ref{table:comp}), with the best fidelity obtained by DE.
We thus enhance DE by introducing modifications into its standard version to enable feasible procedures that lead to intrinsic fidelity beyond 99.99\%. We first give the details of the standard version of DE, and then discuss our machine learning approach, which is called the Subspace-Selective Self-Adaptive Differential Evolution (SuSSADE).

\subsection{Differential Evolution}
Differential evolution is an evolutionary algorithm (EA).
Similar to other EAs, DE is inspired by biological evolution. The robustness and effectiveness of DE have already been studied by researchers in various fields such as machine learning~\cite{Bis06}, optimization~\cite{VT04}, and image classification~\cite{OES05}.

DE is a population-based search heuristic algorithm. Each member of the initial population breeds with the three random members of the same generation to generate a ``daughter''. The fittest of the original member and daughter survive to the next generation.
Three distinct operations exist:
mutation,
crossover and
selection.
These operations form the mathematical structure of DE.
We now explain each of these operations in details.

For each initial population member $D_i$ with $i\in\{1,\dots, P\}$ and $P$ the population size, the mutation operation generates a trail vector $M_i$ as follow:
 \begin{equation}
 \label{eq:mtn}
 M_i=D_{r_{i_1}}+\mu\left(D_{r_{i_2}}-D_{r_{i_3}}\right),
 \end{equation}
where
\begin{equation}
	\{r_{i_1},r_{i_2},r_{i_3}\}\in\{1, \dots, P\}
\end{equation}
are discrete random numbers and mutation rate $\mu$ is a random number uniformly sampled from [0,1]. The mutation rate determines the step-size on the control landscape. A higher value of $\mu$ means that DE tends to explore the un-searched region of the landscape than exploiting the current knowledge about the landscape.

For each initial population member $C_i$ and trial vector $M_i$, the crossover operation generates a target vector $C_i$, such that
\begin{equation}
\label{eq:cr}
	C_i(j)= \begin{cases}
	M_i(j) &\text{if $r_{i_j}<\xi$}\\
	D_i(j) &\text{otherwise},
\end{cases}
\end{equation}
where $j$ is the index denoting the dimension of each population member. In our quantum scheme, the maximum value of $j$ is equal to the number of control parameters $K$. $r_{i_j}$ is a uniform random number sampled from [0,1].~$\xi$ is the crossover rate of the algorithm.
A higher~$\xi$ means that DE exploits the current knowledge about the quantum control landscape without spending too much time in searching the unexplored area of the landscape.

The last operation of DE is the selection operation, where we construct
\begin{equation}
\label{eq:sl}
	D'_i:= \begin{cases}
	C_i &\text{if $f(C_i)>f(D_i)$}\\
	D_i &\text{otherwise},
\end{cases}
\end{equation}
with $f(C_i)$ being any fitness function, which is the fidelity function~(\ref{eq:explicitF}) in our case. In an iterative process, the resultant new population at generation $G$ replaces population in the previous generation $G-1$, and
DE continues with the new generation. The iterative process aborts when either the threshold fidelity reaches or a predefined number of generations is attained. 

Similar to other EAs,
a DE algorithm faces two obstacles when applied to problems with a large number of dimensions. First, finding the optimal algorithmic-parameters i.e., $\mu$ and~$\xi$, which lead to the best performance of DE is computationally expensive. Second, DE converges slowly to the promising region of the landscape where an optimal solution exists. We address these two weakness of DE by proposing the enhanced version, called Subspace-Selective Self-Adaptive DE (SuSSADE).

\subsection{Subspace-Selective Self-Adaptive Differential Evolution}
There are two approaches to find the optimal algorithmic-parameters for DE. One can run DE with many initial guesses to find the optimal parameters. This method is computationally expensive as it needs many trial runs of DE. This method also does not propose a general solution to the problem of finding the algorithmic-parameters because a new set of trial runs is needed if the learning problem is changed.

 An alternative approach is to self-adaptively~\cite{BZM06} change the parameters at each generation $G$ as follows:
\begin{equation}
\label{rate}
	\mu_{i,G+1}= \begin{cases}
	\mu_l+r_1.\mu_u &\text{if $r_2<\kappa_1$}\\
	\mu_{i,G} &\text{otherwise}\\
	\end{cases}
\end{equation}
and
\begin{equation}
\label{rate}
	\xi_{i,G+1}= \begin{cases}
	r_3 &\text{if $r_4<\kappa_2$}\\
	\xi_{i,G} &\text{otherwise,}\\
	\end{cases}
\end{equation}
where
\begin{equation}
	r_j,j\in\{1, 2, 3, 4\}
\end{equation}
are random numbers uniformly sampled from (0,1], and $\mu_l$, $\mu_u$, $\kappa_1$ and $\kappa_2$ are assigned to fixed values of $0.1$, $0.1$, $0.1$ and $0.9$, respectively,
Using a self-adaptive version of DE improved the obtained fidelity up to 0.993,
but the result is still sub-threshold because of the high dimensionality of the learning problem. Therefore, the next step is to enhance
the self-adaptive DE using cooperative coevolution (CC) approach~\cite{PJ94}.

CC decomposes a high-dimensional problem into problems over several subspaces and evolve each sub-component of the original high-dimensional problems using a choice of EA. CC then cooperatively
combines the solution of each sub-component to form the final solution. There are several methods to decompose the high-dimensional problem into smaller sub-components~\cite{PJ94,LYZ+01,YHZ05,SDS02}. The easiest method is to decompose a $K$-dimensional problem
into $K$ one-dimensional problems and evolve each component using EA.

This approach has drawbacks originating from ignoring the interdependencies between variables.
One alternative method is to split the problem dimension in two halves and evolve each half
over the course of learning process. However, should the $K$ be high, the $K/2$ is also high, and therefore, the problem of a large number of dimensions is not addressed properly. Here we take a different approach in decomposing the high-dimensional problem.
Our approach is inspired by Differential Evolution with Cooperative Coevolution (DECC-II)~\cite{ZKX07}. We first give a short description on DECC-II and then explain the enhanced version of DE.

DECC-II decomposes a $K$-dimensional problem into $m$-dimensional subspaces and run the learning algorithm on a subspace for a fixed number of cycles $s$ while keeping the other subspaces unchanged. DECC-II  combines the CC with Self-Adaptive DE with Neighborhood Search (NSDE) to address both the issues: the slow-convergence of DE and finding the optimal algorithmic parameters.

Using DECC-II in its original form did not improve the fidelity and the run-time and convergence of DE still remained slow in our case,
as evaluating the fitness function is computationally expensive.
We also found our result to be influenced by the choice of $s$ and $m$ and finding the optimal values of these parameters needed additional computational resources and trial runs, and even with this overhead the threshold fidelity was not achieved.

Inspired by the DECC-II algorithms, we set $s=1$ and choose the dimension of subspace based on a random projection over the $m$-dimensional subspace with $m\in$\{1, 2, 3, 4, 5\}.
Now, there is no need to look for the optimal values of $s$ and $m$ to perform the optimization. This new strategy, however, makes the convergence slower, as in each generation, only a small part of the candidate solutions are being selected for the optimization.

In order to enhance the convergence, our algorithm randomly switches breeding between the subspace and the whole space
according to the value of an input switch parameter~$S\in[0,1]$,
such that a uniformly distributed random number $r_j\in[0,1]$ at generation~$j$ 
restricts breeding to the subspace, if $r_j<S$, and breeds in the whole space otherwise.
As our algorithm selects an $m$-dimensional subspace at each generation and self-adaptively evolves the mutation and crossover rates, we call it Subspace-Selective Self-Adaptive DE (SuSSADE). For our purpose, we observe that choosing $m=1$ suffices,
which signifies that the selected subspace is trivial. We refer to this one-dimensional extreme case as 1DSuSSADE.

\section{Noise model}
\label{sec:noiseModel}
In order to incorporate decoherence into our system's evolution, we model each transmon as a harmonic oscillator suffering from the environmental effect~\cite{LOM+04}. The decoherence of each harmonic oscillator is represented by two damping rates: amplitude and phase damping with the corresponding amplitude relaxation time $T_1$ and dephasing time $T_2$. In order to make the noise model simpler, we assume $T$=$T_1$=$T_2$, which is a valid assumption for transmons with tunable frequency~\cite{GGZ+13}.

Equation~(\ref{eq:U}) expresses the system evolution in the absence of decoherence. In the presence of noise we model the system evolution by the time-dependent density matrix which is decomposed in terms of the sum of Kraus matrices as ${\mathcal L}_{k}$~\cite{NC05}
\begin{equation}
	\rho(t)=\sum_{k=0}^{n}{\mathcal L}_{k}(t)\rho(0){\mathcal L}^{\dagger}_{k}(t),
\end{equation}
with Kraus matrices satisfying the completeness relation
\begin{equation}
\label{eq:kraus}
\sum_{k=0}^{n}{\mathcal L}^{\dagger}_{k}{\mathcal L}_{k}=\mathds{1}
\end{equation}
at each time step. We first discuss the construction of the Kraus matrices for each transmon.

Constructing the Kraus matrix representation for a system comprising three capacitively coupled transmons, will then be straightforward by performing all the possible tensor products of three transmons. We also assume that the decoherence only affect each individual transmon separately.

\subsection{Amplitude damping}
The Kraus matrix representation of amplitude damping of a single qubit (treated as a truncated harmonic oscillator) coupled to the environment
can  be modeled as multimode oscillators~\cite{LOM+04}.
One can easily generalize this approach to represent the
Kraus representation of amplitude damping of a single transmon (modeled as a four-modal harmonic oscillator) coupled to environment modeled as multimode oscillators, where
\begin{align}
\label{eq:Am_damping}
	A_l(t)=\sum_{j=l}^{3}\sqrt{j \choose l}&\left(\text{e}^{-\frac{t}{T_{1}}}\right)^{\frac{j-l}{2}}\nonumber\\&
		\left(1-\text{e}^{-\frac{t}{T_{1}}}\right)^{\frac{l}{2}}\ket{j-l}\bra{j}
\end{align}
with $l\in$\{0, 1, 2, 3\} labeling the Kraus matrices
and $\text{exp}(-t/T)$ represents the amplitude damping factor,
which decays exponentially with the timescale $T_1$.

\subsection{Phase damping}
We express the Kraus matrix representation of the phase damping of a single transmon (with four energy levels) coupled to the environment~\cite{LOM+04} as
\begin{equation}
\label{eq:Ph_damping}
	{\mathbb A}_l(t)=\sum_{j=0}^{3}\exp\left\{-\frac{j^{2}t}{2T_\text{2}}\right\}\sqrt{\frac{(j^{2}t/T_\text{2})^{l}}{l!}}\ket{j}\bra{j}
\end{equation}
with $l\in\{1, 2, 3, 4\}$. ${\mathbb A}_l$ is a Kraus matrix,
and $T_2$ indicates the timescale for dephasing.
In both equations~(\ref{eq:Am_damping}) and~(\ref{eq:Ph_damping}), we need to put an upper bound on the Kraus-Matrix index $l$ to enable numerical simulation of decoherence.

Such an upper bound would violate the
completeness relation~(\ref{eq:kraus}). However, if the evolution time of the quantum system $t$ is much smaller than the
coherence time $T$, the higher-order terms in~(\ref{eq:Am_damping}) and~(\ref{eq:Ph_damping}) damp exponentially with respect to the $t/T$. As in our case $t\ll{T}$, we only consider $l$ up to three for both amplitude and phase damping, and discard the higher-order terms in our numerical calculation as they have negligible effects.

\section{Three-qubit logical gates}
\label{sec:threeQubitLogicalGates}
In this section we give the details of the Toffoli, Fredkin and CXX gates. We discuss the circuit model of each gate as well as give the truth tables showing how the basis states transform under the actions of these gates.

\subsection{Toffoli  gate}

A Toffoli gate is a three-qubit gate that applies a Pauli X operation on the third qubit if the quantum state of the first two control qubits are $\ket{11}$, and does nothing otherwise. A Toffoli gate is also called a Controlled-Controlled-NOT (CCX) gate,
which is equivalent to a Controlled-Controlled-Z (CCZ)
gate up to a local Hadamard transformation on the target qubit~\cite{ZGS15}.

CCZ gate is a three qubit gate that applies a Pauli-Z operator on the third qubit if the quantum state of the first two qubits are $\ket{11}$ and, otherwise leaves the state of the third qubit unchanged.
%Fig.~\ref{fig:Toffoli_CZ} demonstrates the circuit representation of Toffoli gate and its equivalent CCZ gate.
In order to design the Toffoli gate, we only show how to implement the CCZ gate, since the Hadamard gates are trivial for superconducting circuits~\cite{KBC+14,MGR+09}.

In our supervised learning method,
the truth table of the CCZ gate represents the training set. One can define the truth table of the CCZ gate based on its action on the three-qubit basis states (See Table~\ref{table:CCZtruth}).
\begin{table}[t]
\centering
\begin{tabular}{c|c|c|c|c|c}
     \multicolumn{3}{c|}{Input} &
    \multicolumn{3}{c}{Output} \\ \hline
	$C_1$ &$C_2$& $T$ & $C_1$& $C_2$& $T$    \\\hline
	$\ket{0}$  &$\ket{0}$ &$\ket{0}$ &$\ket{0}$&$\ket{0}$&$\ket{0}$  \\
	$\ket{0}$  &$\ket{0}$ &$\ket{1}$ &$\ket{0}$&$\ket{0}$&$\ket{1}$  \\
	$\ket{0}$  &$\ket{1}$ &$\ket{0}$ &$\ket{0}$&$\ket{1}$&$\ket{0}$  \\
	$\ket{0}$  &$\ket{1}$ &$\ket{1}$ &$\ket{0}$&$\ket{1}$&$\ket{1}$  \\
	$\ket{1}$  &$\ket{0}$ &$\ket{0}$ &$\ket{1}$&$\ket{0}$&$\ket{0}$  \\
	$\ket{1}$  &$\ket{0}$ &$\ket{1}$ &$\ket{1}$&$\ket{0}$&$\ket{1}$  \\
	$\ket{1}$  &$\ket{1}$ &$\ket{0}$ &$\ket{1}$&$\ket{1}$&$\ket{0}$  \\
	$\ket{1}$  &$\ket{1}$ &$\ket{1}$ &$\ket{1}$&$\ket{1}$&-$\ket{1}$  \\
\end{tabular}
\caption{The truth table for CCZ gate. $C_1$ and $C_2$ denote the control qubits and $T$ represents the target qubit.%
}
\label{table:CCZtruth}
\end{table}

So far, the design of a high-fidelity Toffoli gate has been investigated in several physical systems with the achieved fidelities limited to 81\% in a post-selected photonics
circuit~\cite{LBA+09}, 71\% in an ion trap system~\cite{MKH+09}, 68.5\% in a three-qubit circuit QED~\cite{FSB+12} and 78\% in a four-qubit circuit QED~\cite{RDN+12}. We recently proposed a quantum control
approach to design a high-fidelity ($>$99.9\%) Toffoli gate for a system comprising three nearest-neighbor-coupled transmons~\cite{ZGS15}.

 \subsection{Fredkin gate}
 \begin{figure}
	\includegraphics[width=.6\columnwidth]{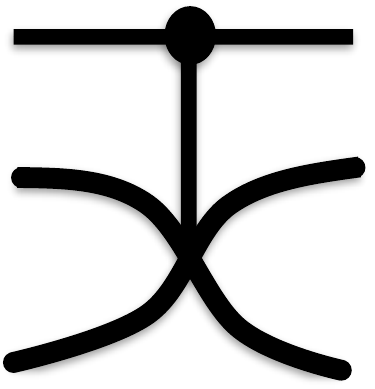}
\caption{The quantum circuit representation of Fredkin (Controlled-Swap) gate. The horizontal solid back line is the circuit wire, {\large${\bullet}$} denotes the control qubit and the big cross sign shows the SWAP gate which acts on the target qubits (second and third qubits).}
\label{fig:Fredkin}
\end{figure}
 
A Fredkin gate (See Fig.~\ref{fig:Fredkin}) is a three-qubit gate that applies a SWAP operation between the second and the third qubits, if the state of the first qubit is~$\ket{1}$, and leaves the state of the qubits unchanged otherwise. A Fredkin gate is an excitation-number-preserving operation,
 where the output state has the same number of excitations as that for the input states. It is also a self-inverse operation, which means that applying two consecutive Fredkin gates generates the Identity operation.
 The Fredkin gate is universal for reversible classical computation, as it can be used to construct any other reversible classical logic gates~\cite{FT82}.
 
The truth table of Fredkin gate is shown in Table~\ref{table:Fredtruth}. In this representation $C$ denotes the control qubit and $T_1$ and $T_2$ represent the target qubits. This truth table provides the training set for Fredkin for supervised learning. 
\begin{table}[t]
\centering
\begin{tabular}{c|c|c|c|c|c}
     \multicolumn{3}{c|}{Input} &
    \multicolumn{3}{c}{Output} \\ \hline
	$C$ &$T_1$& $T_2$ & $C$& $T_1$& $T_2$    \\\hline
	$\ket{0}$  &$\ket{0}$ &$\ket{0}$ &$\ket{0}$&$\ket{0}$&$\ket{0}$  \\
	$\ket{0}$  &$\ket{0}$ &$\ket{1}$ &$\ket{0}$&$\ket{0}$&$\ket{1}$  \\
	$\ket{0}$  &$\ket{1}$ &$\ket{0}$ &$\ket{0}$&$\ket{1}$&$\ket{0}$  \\
	$\ket{0}$  &$\ket{1}$ &$\ket{1}$ &$\ket{0}$&$\ket{1}$&$\ket{1}$  \\
	$\ket{1}$  &$\ket{0}$ &$\ket{0}$ &$\ket{1}$&$\ket{0}$&$\ket{0}$  \\
	$\ket{1}$  &$\ket{0}$ &$\ket{1}$ &$\ket{1}$&$\ket{1}$&$\ket{0}$  \\
	$\ket{1}$  &$\ket{1}$ &$\ket{0}$ &$\ket{1}$&$\ket{0}$&$\ket{1}$  \\
	$\ket{1}$  &$\ket{1}$ &$\ket{1}$ &$\ket{1}$&$\ket{1}$&$\ket{1}$  \\
\end{tabular}
\caption{The truth table representation of Fredkin gate. $C$ denotes the control qubit and $T_1$ and $T_2$ represent the target qubit. The column under the Output and Input columns show the state of the three qubits before and after applying the Fredkin gate.}
\label{table:Fredtruth}
\end{table}

Proposals to implement the Fredkin gate are mainly restricted to the context of linear and nonlinear optical systems, and there has been no proposals yet for a single-shot Fredkin gate with superconducting circuits. 
Here we employ the quantum control scheme to design a fast single-shot high-fidelity Fredkin gate for a system comprising three nearest-neighbor-coupled superconducting transmon systems.

\subsection{CXX}
CXX is a three-qubit gate that applies Pauli-X operations on the second and third qubits when the first qubit is~$\ket{1}$, and leaves the state of the qubits unaltered otherwise. CXX gate is equivalent to  Controlled-Z-Z (CZZ) gate up to local Hadamard operations on both the second and the third qubits:
\begin{equation}
\label{eq:CZZ}
	\text{CXX}=\left[\mathds{1}{\otimes}H\otimes H\right]
	\text{CZZ}\left[\mathds{1}{\otimes}H\otimes H\right].
\end{equation}
 The CZZ gate is a three-qubit gate that applies a Pauli-Z operation on the second and the third qubits, if and only if, the first qubit is~$\ket{1}$. In order to design the CXX gate, we also assume that the implementation of fast and high-fidelity Hadamard gates are trivial for superconducting circuits~\cite{KBC+14,MGR+09};
hence we employ the quantum control scheme to design the high-fidelity CZZ gate.

 CZZ gate can easily be constructed by consecutive operations of two CZ gates. A high-fidelity CZ gate has already been proposed with the gate time about 26 \text{ns}~\cite{GGZ+13}. Therefore, a high-fidelity CZZ gate decomposed into two consecutive CZ gates can be readily designed with a gate operation time no longer than 52 \text{ns}. However, a CZZ gate with a shorter operation time is more useful for quantum error correction, which motivates us to employ the quantum control scheme for this problem.

The truth table representation of a CZZ gate is shown in Table~\ref{table:CZZtruth}. In this table, $C$ denotes the control qubit and $T_1$ and $T_2$ represent the target qubits. We use the truth table data as the training set to train the qubit frequencies (our hypothesis).
\begin{table}[t]
\centering
\begin{tabular}{c|c|c|c|c|c}
     \multicolumn{3}{c|}{Input} &
    \multicolumn{3}{c}{Output} \\ \hline
	$C$ &$T_1$& $T_2$ & $C$& $T_1$& $T_2$    \\\hline
	$\ket{0}$  &$\ket{0}$ &$\ket{0}$ &$\ket{0}$&$\ket{0}$&$\ket{0}$  \\
	$\ket{0}$  &$\ket{0}$ &$\ket{1}$ &$\ket{0}$&$\ket{0}$&$\ket{1}$  \\
	$\ket{0}$  &$\ket{1}$ &$\ket{0}$ &$\ket{0}$&$\ket{1}$&$\ket{0}$  \\
	$\ket{0}$  &$\ket{1}$ &$\ket{1}$ &$\ket{0}$&$\ket{1}$&$\ket{1}$  \\
	$\ket{1}$  &$\ket{0}$ &$\ket{0}$ &$\ket{1}$&$\ket{0}$&$\ket{0}$  \\
	$\ket{1}$  &$\ket{0}$ &$\ket{1}$ &$\ket{1}$&$\ket{0}$&-$\ket{1}$  \\
	$\ket{1}$  &$\ket{1}$ &$\ket{0}$ &$\ket{1}$&-$\ket{1}$&$\ket{0}$  \\
	$\ket{1}$  &$\ket{1}$ &$\ket{1}$ &$\ket{1}$&$\ket{1}$&$\ket{1}$  \\
\end{tabular}
\caption{The truth table representation of CZZ gate. $C$ denotes the control qubit and $T_1$ and $T_2$ represent the target qubits.
The columns under the Output and Input show the states of the three qubits before and after applying CZZ.}
\label{table:CZZtruth}
\end{table}

\begin{figure}
\includegraphics[width=\columnwidth]{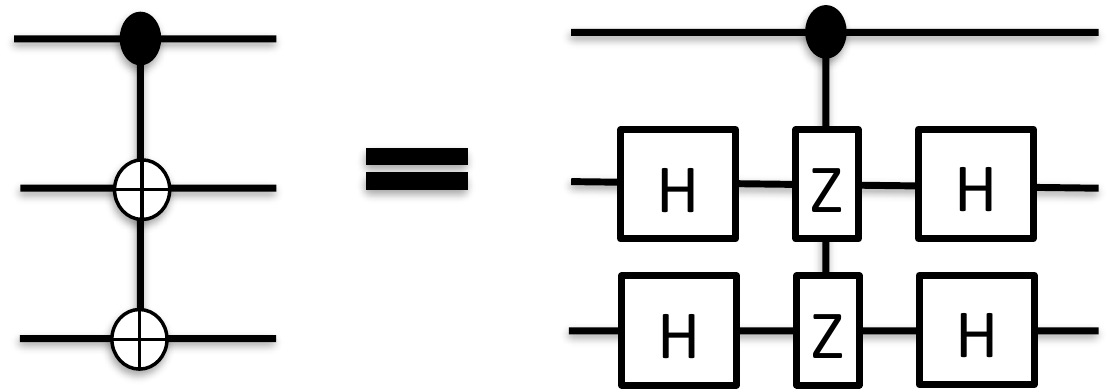}
\caption{The quantum circuit representation of the CXX gate (left), which is equivalent to the CZZ gate up to local Hadamard gates (right). The horizontal solid black lines are circuit wires, {\large${\bullet}$} represents the control qubit and $\bigoplus$ denotes the Pauli-X operator acting on the target qubit. The boxes with Z and H denote the Pauli-Z and Hadamard operations, respectively.}
\label{fig:CXX}
\end{figure}

\section{results}
\label{sec:result}
In this section, we first review the results for designing a high-fidelity single-shot Toffoli gate from Ref.~\cite{ZGS15}, and then present our analysis of performance of the Toffoli gate against the effect of random noise on the learning parameters (not given in Ref.~\cite{ZGS15}). Next we present the results for designing single-shot high-fidelity Fredkin and CXX gates. For each of these gates, we first give the optimal piecewise-constant and piecewise-error-function pulses that steer the system evolution towards the threshold fidelity. We then investigate the dependence of intrinsic fidelities on the coupling strength between transmons.

We test robustness of our optimal pulses for Fredkin and CXX gates by applying
uniformly distributed random noise on each of the learning parameters and then calculating the intrinsic fidelity in the presence of such noise. Finally, we give the result for the average state fidelities in the presence of decoherence induced noise.

\subsection{Toffoli}
A set of piecewise-constant and piecewise-error-function pulses are obtained via the SuSSADE algorithm for a single-shot high-fidelity Toffoli gate, which operates over 26~ns (See Fig.~1 in~\cite{ZGS15}) and has the same number of learning parameters for both pulses. The resultant fidelity for the Toffoli gate is higher than $0.999$, even in the presence of decoherence (See Fig.~3 in Ref.~\cite{ZGS15}).

We have also explored the dependence of the intrinsic fidelity over the time of the system evolution by fixing the coupling strength~$g$ to various values and running the SuSSADE for the less-computationally-expensive piecewise-constant pulses. The control pulses are discretized into equally-spaced time intervals of $1~\text{ns}$, which give rise to learning problems with different learning parameters.  For each value of~$g$ we have also shown the dependence of intrinsic fidelity over the gate operation time (See Fig.~2A in Ref.~\cite{ZGS15}).

\begin{figure}
	\includegraphics[width=\columnwidth]{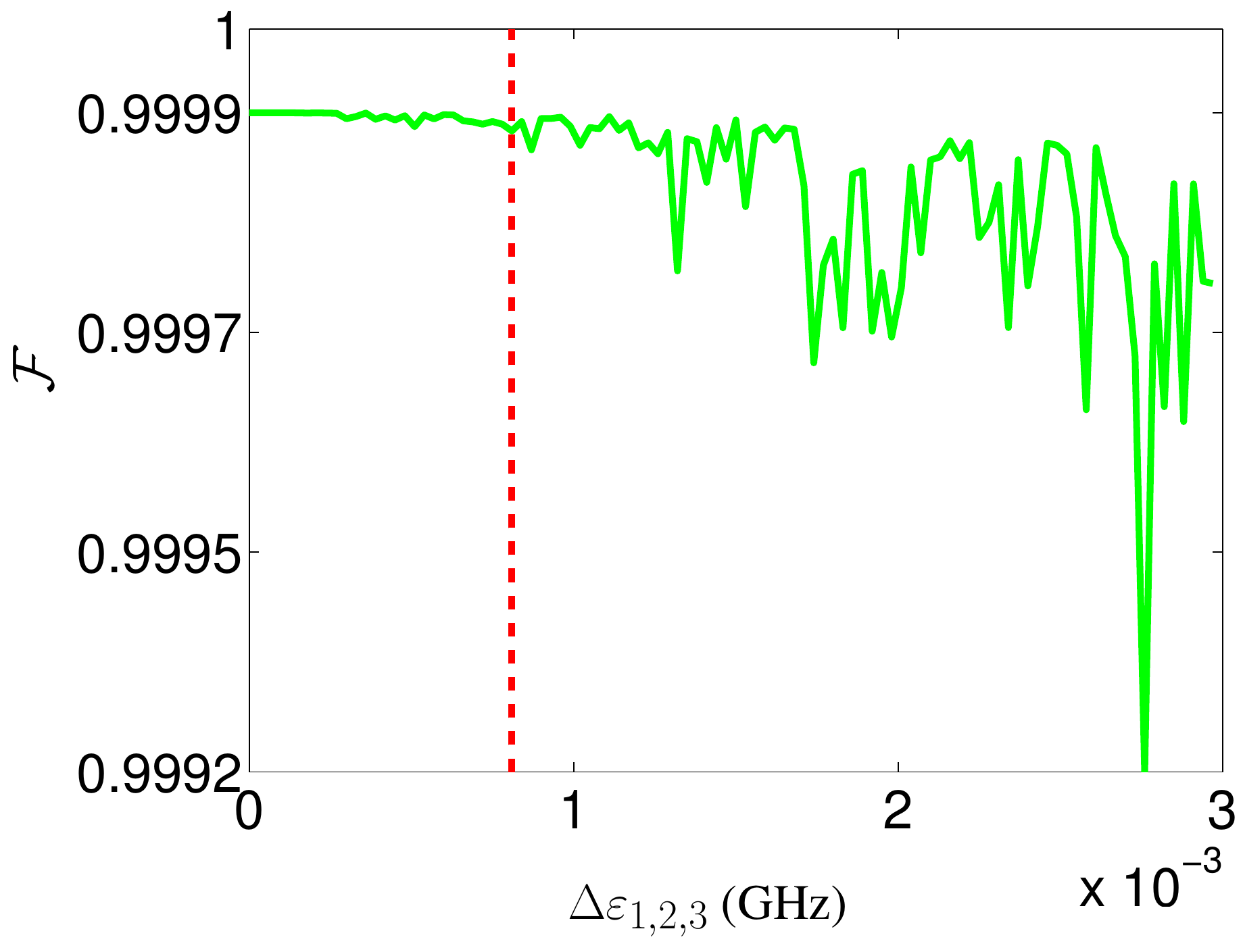}
\caption{(color online) Intrinsic fidelity of the Toffoli gate $\mathcal{F}$ as a function of $\delta\varepsilon$ for the CCZ gate.
The vertical red dotted-line denotes the threshold, such that $\mathcal{F}>0.9999$ on the left of the dotted line.}
\label{fig:T_robustness}
\end{figure}

We study the robustness of the designed pulse against the random noise on the learning parameters. In order to test the robustness we choose a sample optimal pulse for the Toffoli and add random values ($\delta\varepsilon\times{\text{rand(-1,1)}}$) to the learning parameters at each time-bin, with $\delta\varepsilon$ varies from $0$ to $3000$~KHz. We then use the distorted pulse to calculate the intrinsic fidelity for each value of $\delta\varepsilon$. Figure~\ref{fig:T_robustness} shows the change in intrinsic fidelity originated from such random noise on learning parameters.

\subsection{Fredkin}
\begin{figure}
	\includegraphics[width=\columnwidth]{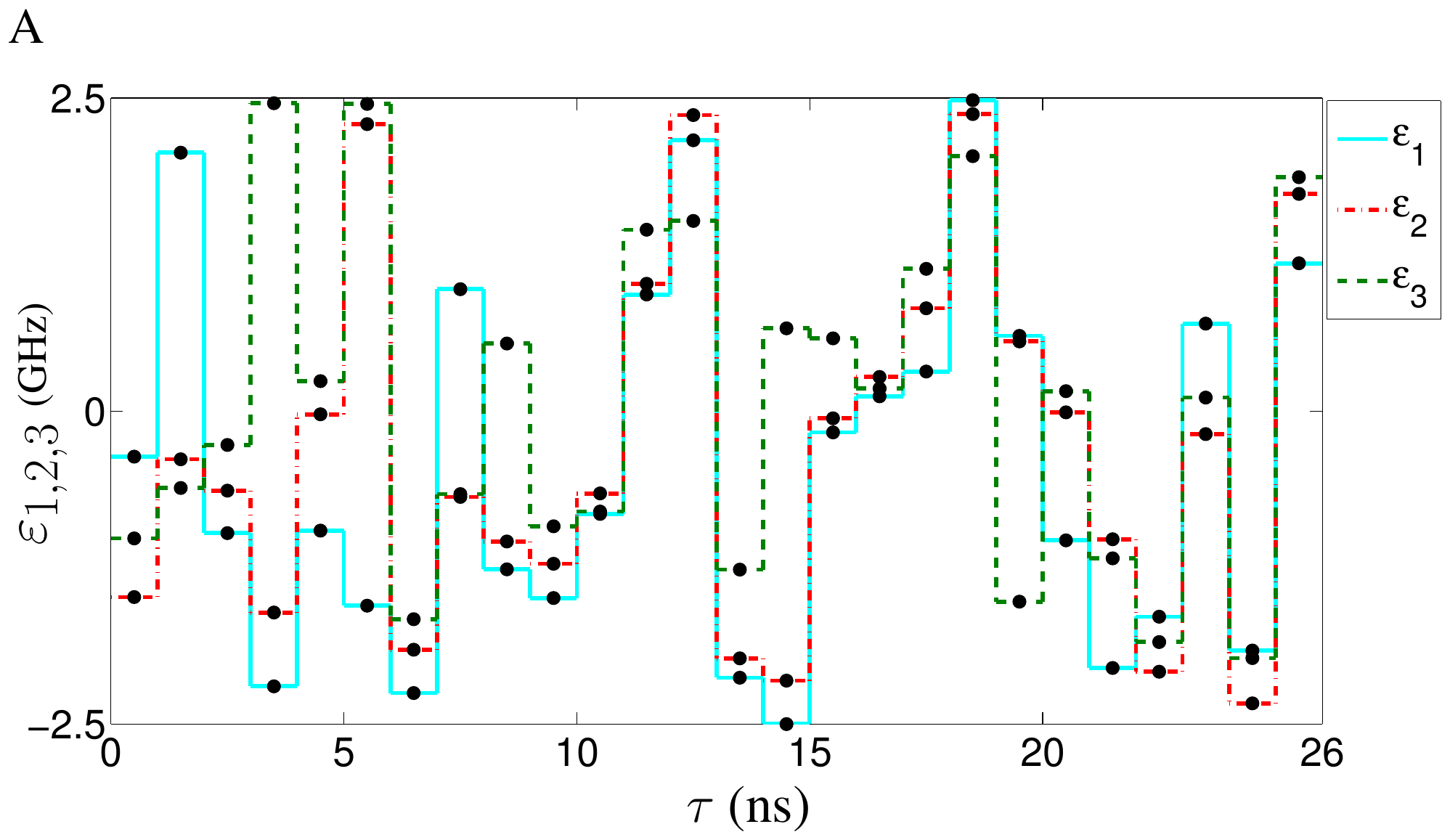}
	\includegraphics[width=\columnwidth]{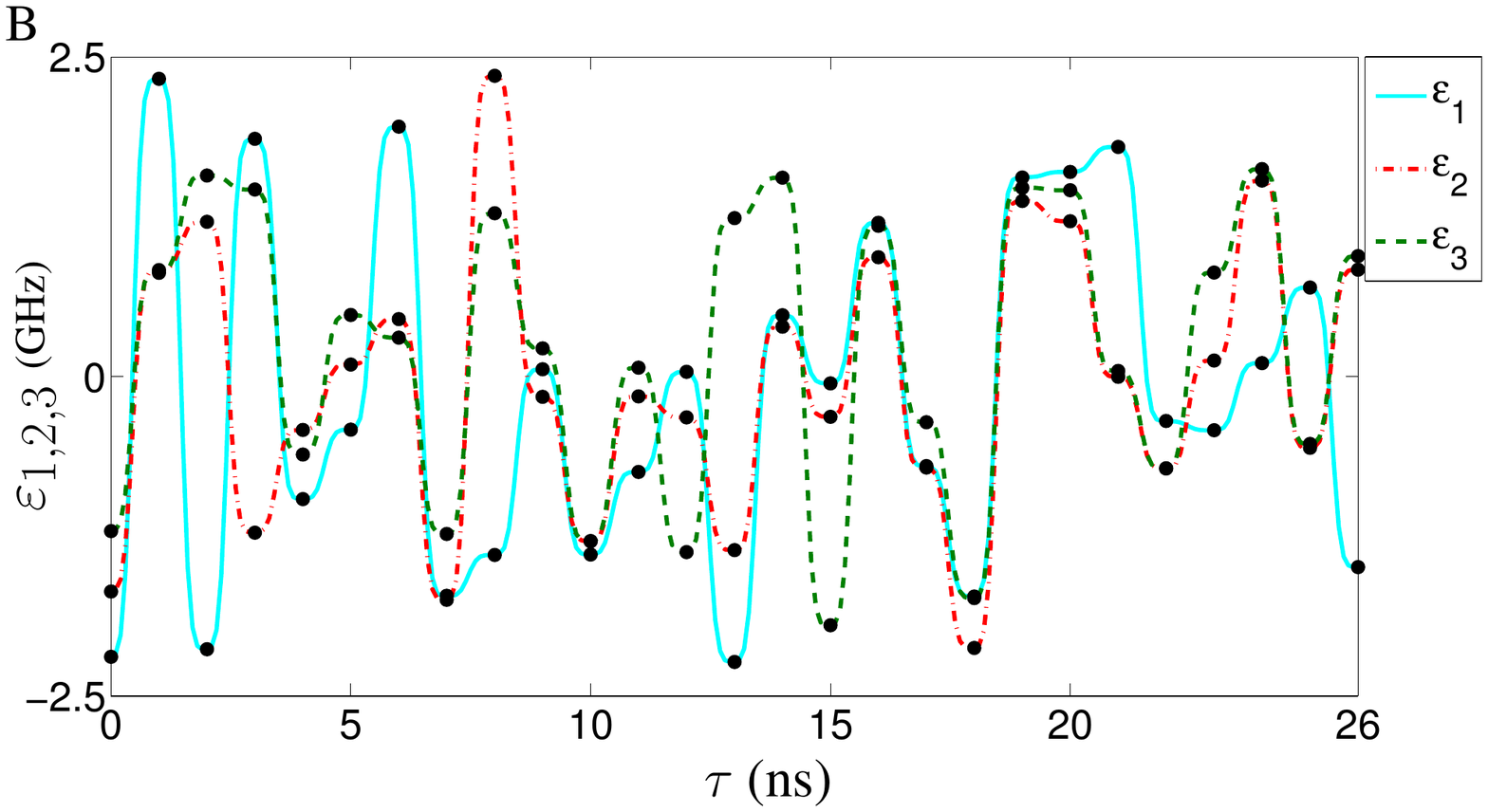}
\caption{
	(color online) Optimal pulses for designing Fredkin gate with the resultant fidelity better than $0.999$ and the gate operation time of $26~\text{ns}$.
	System frequencies, $\varepsilon_i$, are varied from -2.5 to 2.5~GHz, which are within the experimental requirements of transmon implementation. The black dots denote the learning parameters for SuSSADE.
	A) The piecewise-constant pulses for each transmon frequency.
	B) The piecewise-error-function pulses for each transmon frequency.
	}
\label{fig:F_recp}
\end{figure}

We employ SuSSADE to design a high-fidelity Fredkin gate for a system comprising three nearest-neighbor-coupled superconducting transmons. The system Hamiltonian evolves over $26$~\text{ns} under the piecewise-constant pulse (See Fig.~\ref{fig:F_recp}A), where the final evolution approximates a Fredkin gate with an intrinsic fidelity of $\mathcal{F}=0.9999$. We also show a more realistic piecewise-error-function pulse in Fig.~\ref{fig:F_recp}B. The learning algorithm uses the same number of learning parameters to shape the transmon
frequencies through the learning procedure.

\begin{figure}
	\includegraphics[width=\columnwidth]{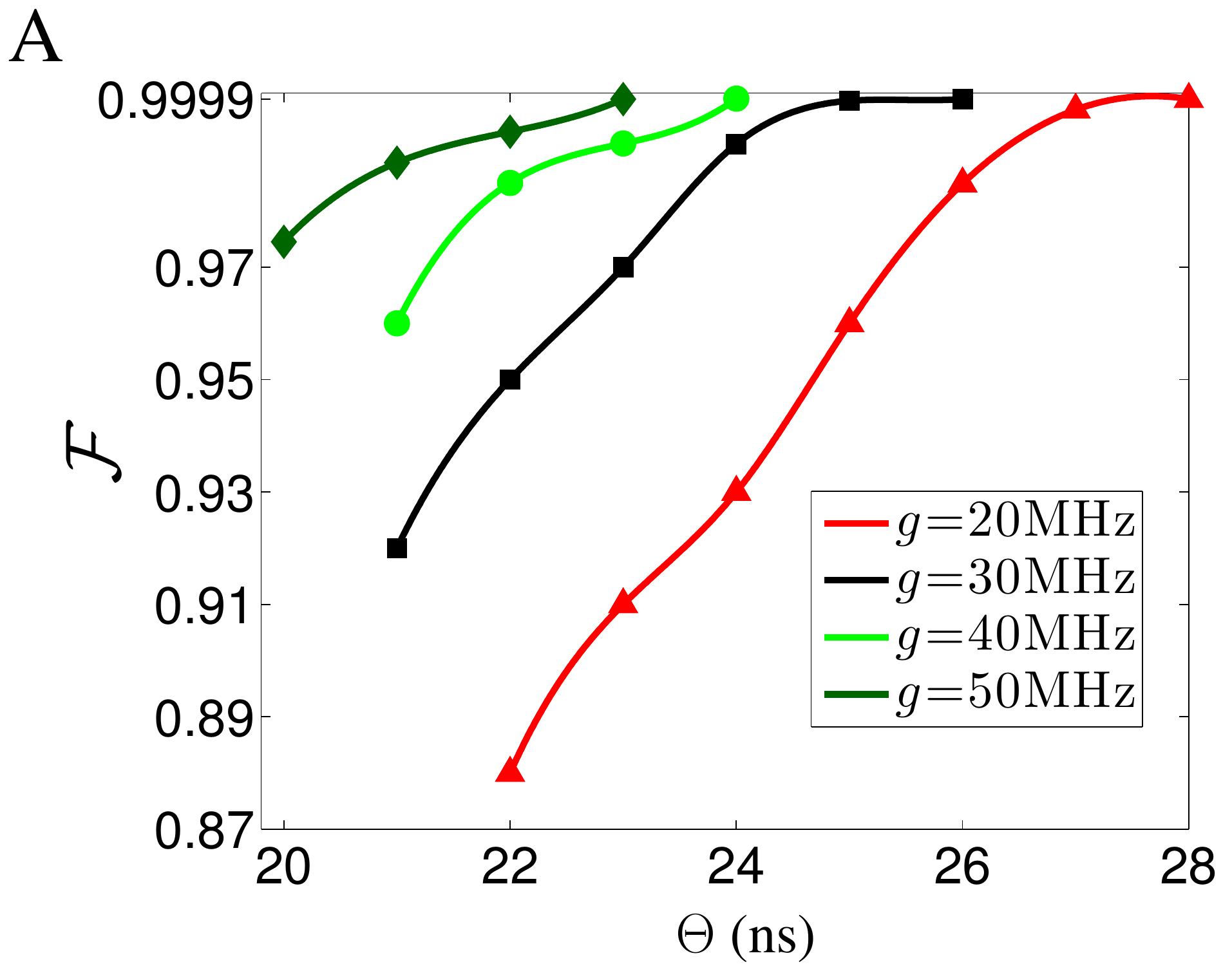}
	\includegraphics[width=\columnwidth]{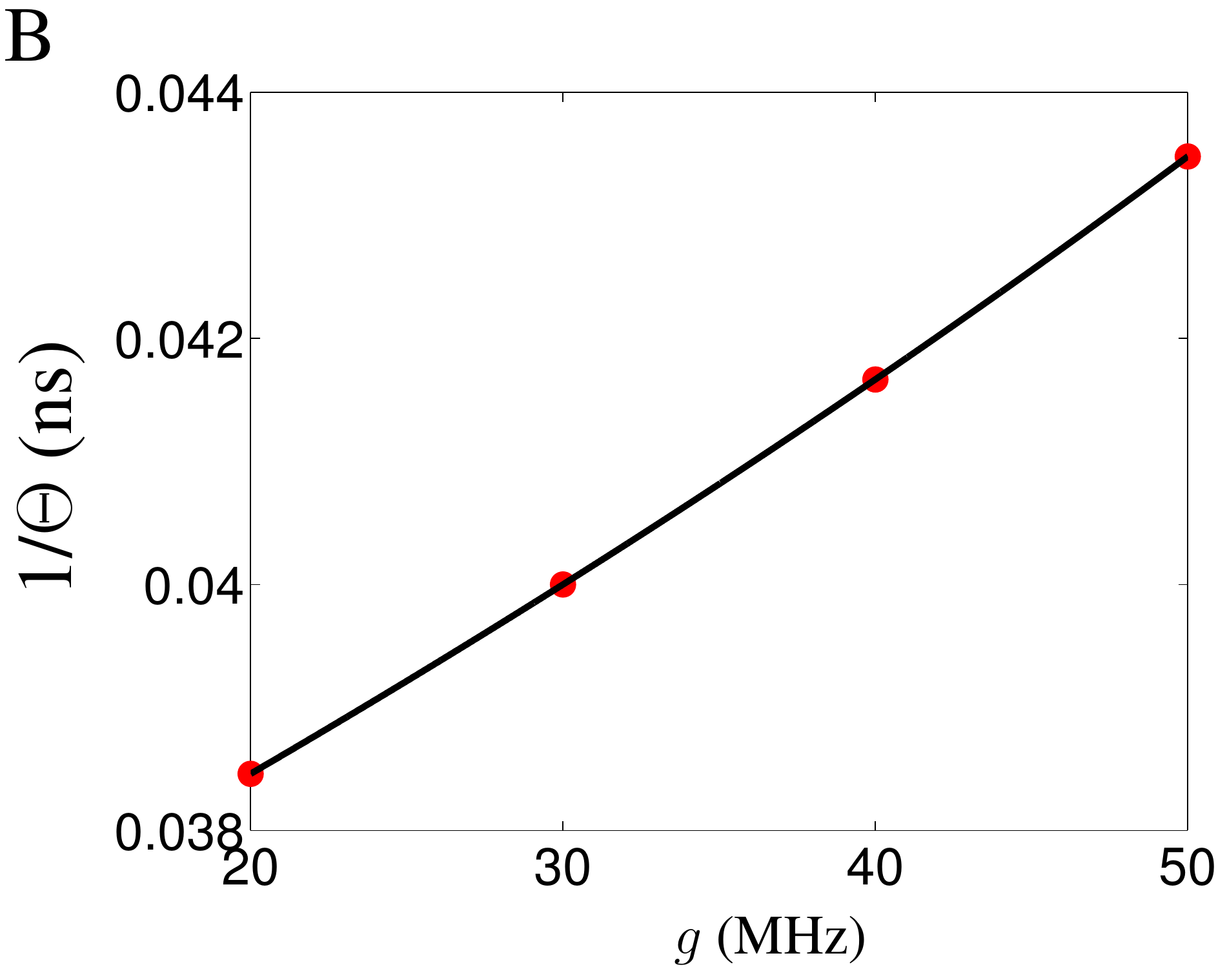}
\caption{
	(color online) A) The dependence of intrinsic fidelity of the Fredkin gate on the evolution time $\tau$ of the system for various values of~$g$. The discretized values show the actual numerical data with $\triangle$, $\square$, $\circ$, $\Diamond$ corresponding to
	the values of~$g$ to be 20, 30, 40, 50~\text{MHz}, respectively. A cubic interpolation fits the curves to the data. B) The relation between the inverse of the gate operation time and coupling strength between transmons where the dots
	denote the actual numerical results for various values of $g\in\{20, 30, 40, 50\}$. A linear-fit interpolates the points to the actual data.
	}
\label{fig:F_fidvst}
\end{figure}

In order to show that the efficacy of the quantum control approach does not depend on the type of the gate, we conduct the same
analyses on the Fredkin gate as we did for Toffoli. In Fig.~\ref{fig:F_fidvst}A we analyze the change in intrinsic fidelity with the gate operation time
for different values of coupling strengths. In Fig.~\ref{fig:F_fidvst}B we fix the intrinsic fidelity at $\mathcal{F}=0.999$ and compute the relation between
coupling strength and the inverse of the gate operation time, where the points on the curve show the actual numerical data, and the solid
line is the cubic-fitted interpolation plot.

\begin{figure}
	\includegraphics[width=0.9\columnwidth]{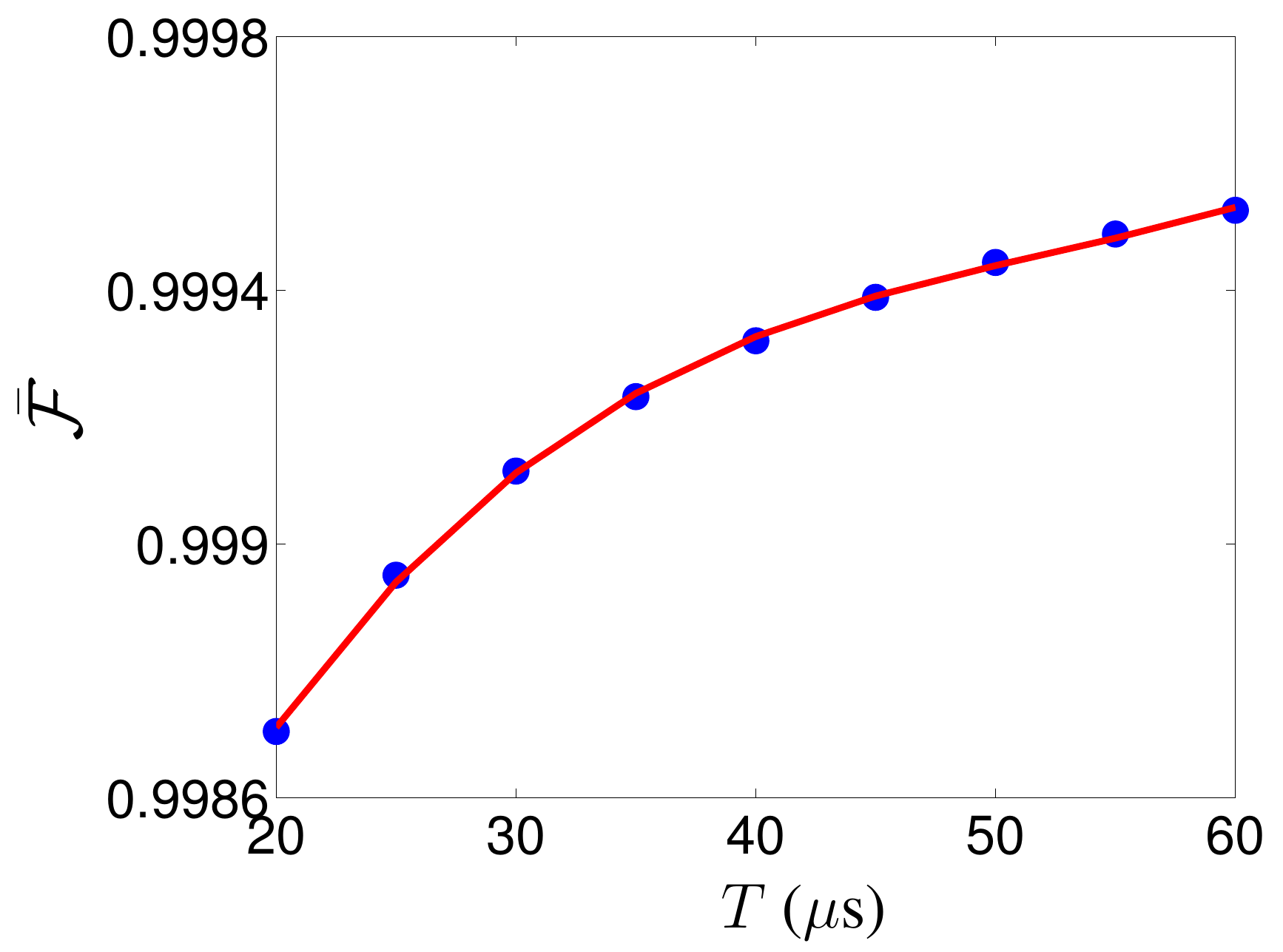}
\caption{
	(color online)
	Fidelity vs coherence time for the Fredkin gate. The dots denote the actual numerical data and the red solid line shows a cubic-fit interpolation on the actual data.}
\label{fig:F_decoherence}
\end{figure}

Figure~\ref{fig:F_decoherence} shows the evolution of the system under the decoherence. We set $g=30~\text{MHz}$ and evolve the system towards the Fredkin gate with an intrinsic fidelity higher than $\mathcal{F}=0.9999$.
Then we apply the noise model on each transmon to analyze how the fidelity changes over the coherence time, $T$, of each transmon. Under our noise model each transmon goes under amplitude- and phase-damping, and we assume $T=T_1=T_2$ for tunable transmons.

\begin{figure}
	\includegraphics[width=0.9\columnwidth]{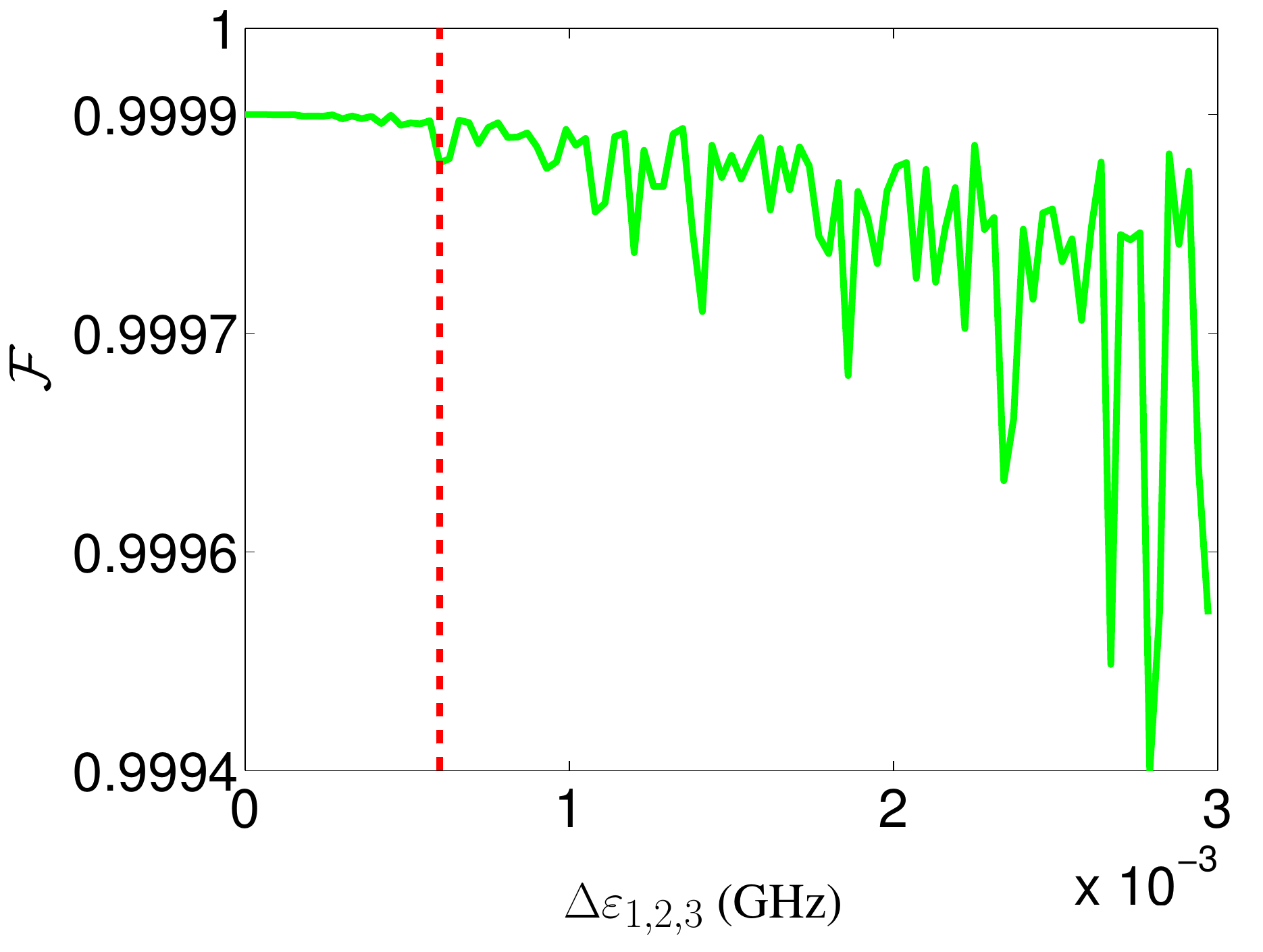}
\caption{Intrinsic fidelity $\mathcal{F}$ vs random noise applied to the optimal pulse of the Fredkin gate. The vertical red dotted-line denotes the threshold, such that on the left side of the line $\mathcal{F}>0.9999$.}
\label{fig:F_robustness}
\end{figure}

Figure~\ref{fig:F_robustness} shows the effect of random noise on the learning parameters of the Fredkin gate. We choose the optimal pulse shown in Fig.\ref{fig:F_recp} and apply random noise up to 3000~KHz on the learning parameters.
We then use the distorted pulse to investigate the change in the intrinsic fidelity as a function of the applied random noise.

\subsection{Controlled-Z-Z (CZZ)}
\begin{figure}
	\includegraphics[width=0.9\columnwidth]{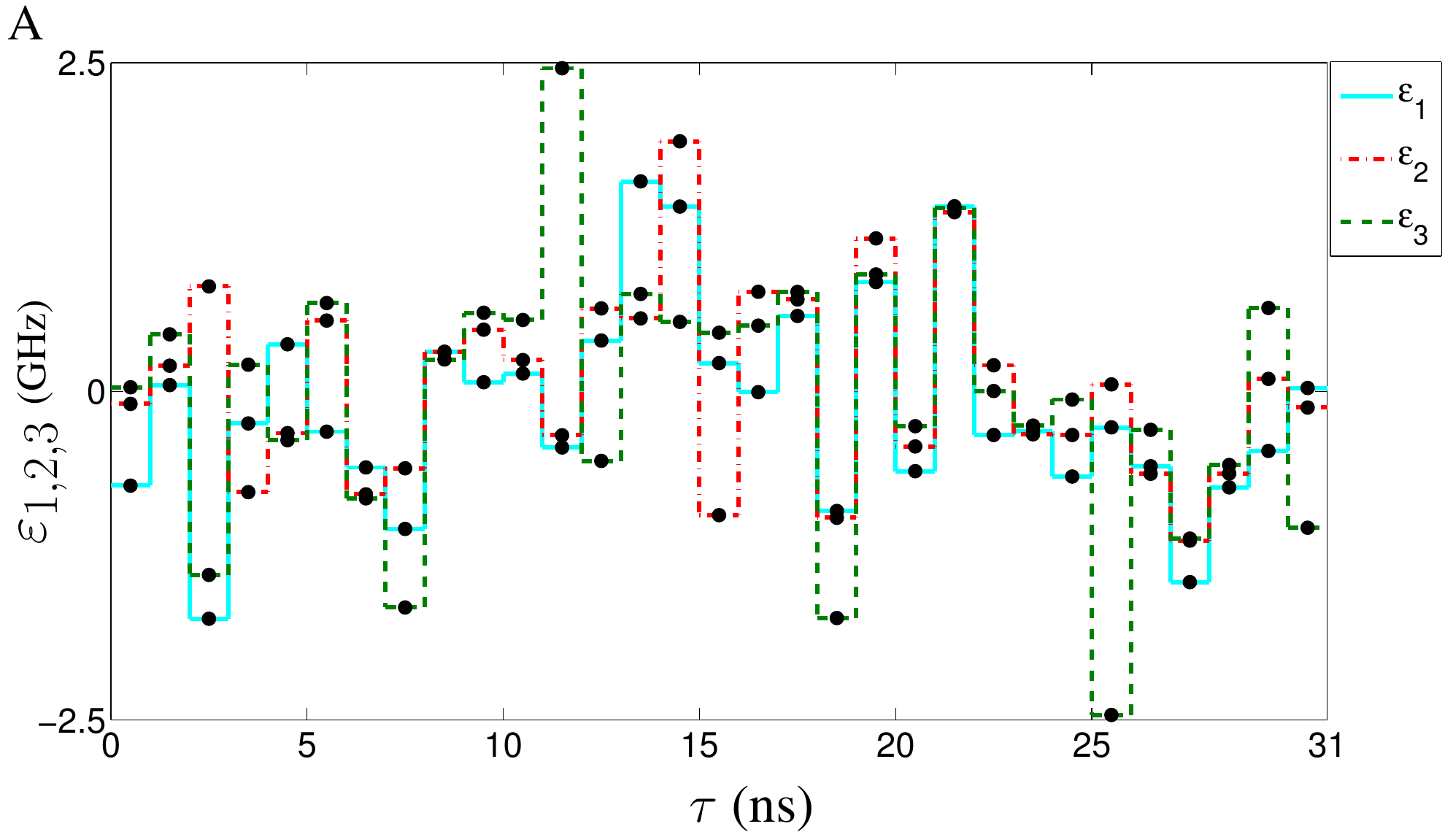}
	\includegraphics[width=0.9\columnwidth]{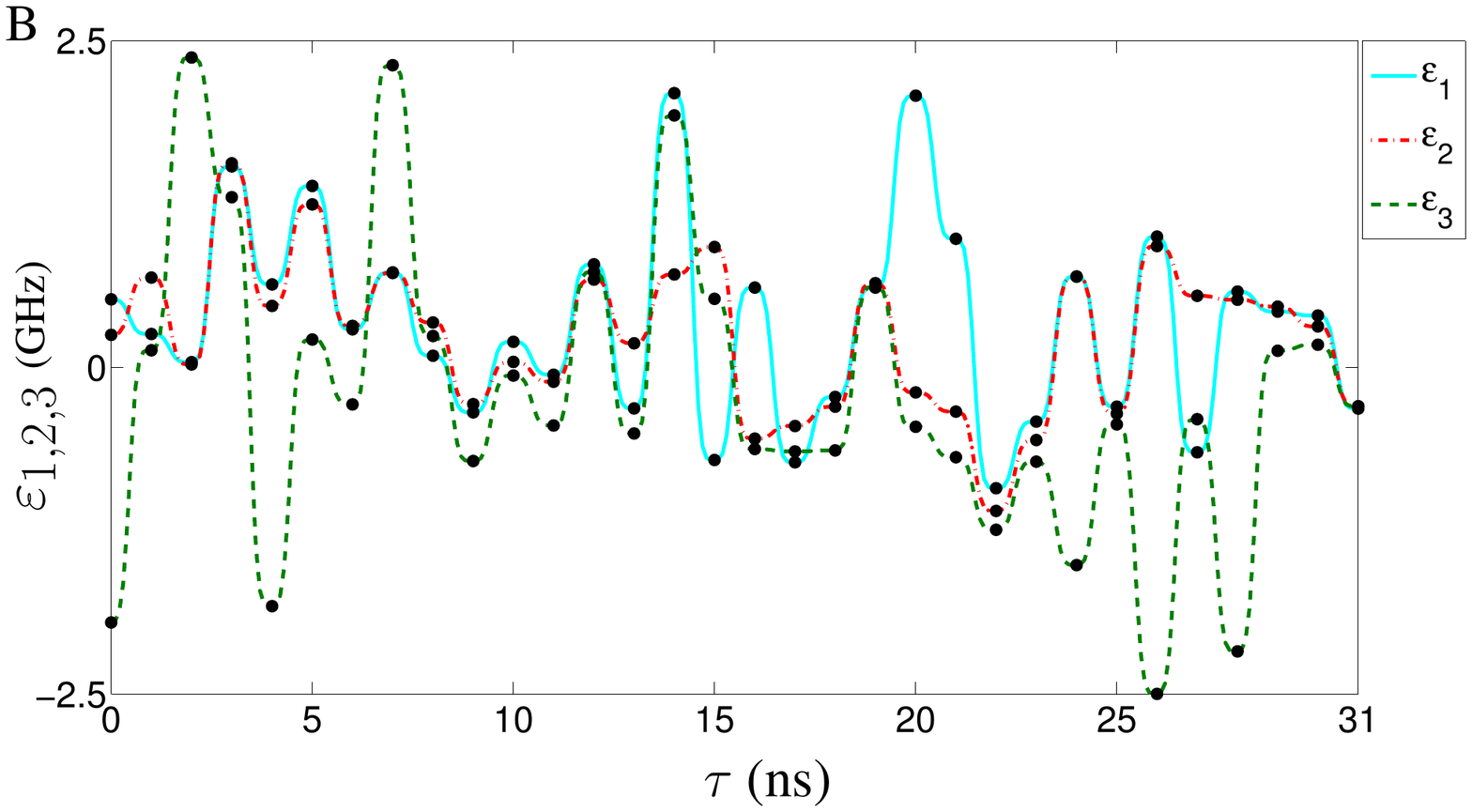}
\caption{
	(color online) Optimal pulses for designing CZZ gate with the resultant fidelity better than $0.999$ and the gate operation time of $31~\text{ns}$.
	System frequencies, $\varepsilon_i$, vary from -2.5 to 2.5~GHz which are within the experimental constraints of transmon implementation.
	The black dots denote the learning parameter for SuSSADE.
	A) The piecewise-constant pulses for each transmon frequency.
	B) The piecewise-error-function pulses for each transmon frequency.
	}
\label{fig:C_recp}
\end{figure}

Fig.~\ref{fig:C_recp}A shows the piecewise-constant pulse generating a high-fidelity CZZ gate in $31$~\text{ns} with an intrinsic fidelity higher than $0.9999$. We have $31$ learning parameters for each pulse ($93$ in total)
to design the pulse shapes for the transmon frequencies. Under the piecewise-error-function pulse (shown in Fig.~\ref{fig:C_recp}B), the system evolution approximate CZZ gate with a fidelity higher than $0.9999$ in the same gate operation time as the
piecewise-constant pulse. In designing the optimal pulses in Fig.~\ref{fig:C_recp} we set $g=30~\text{MHz}$.

\begin{figure}
	\includegraphics[width=0.49\columnwidth]{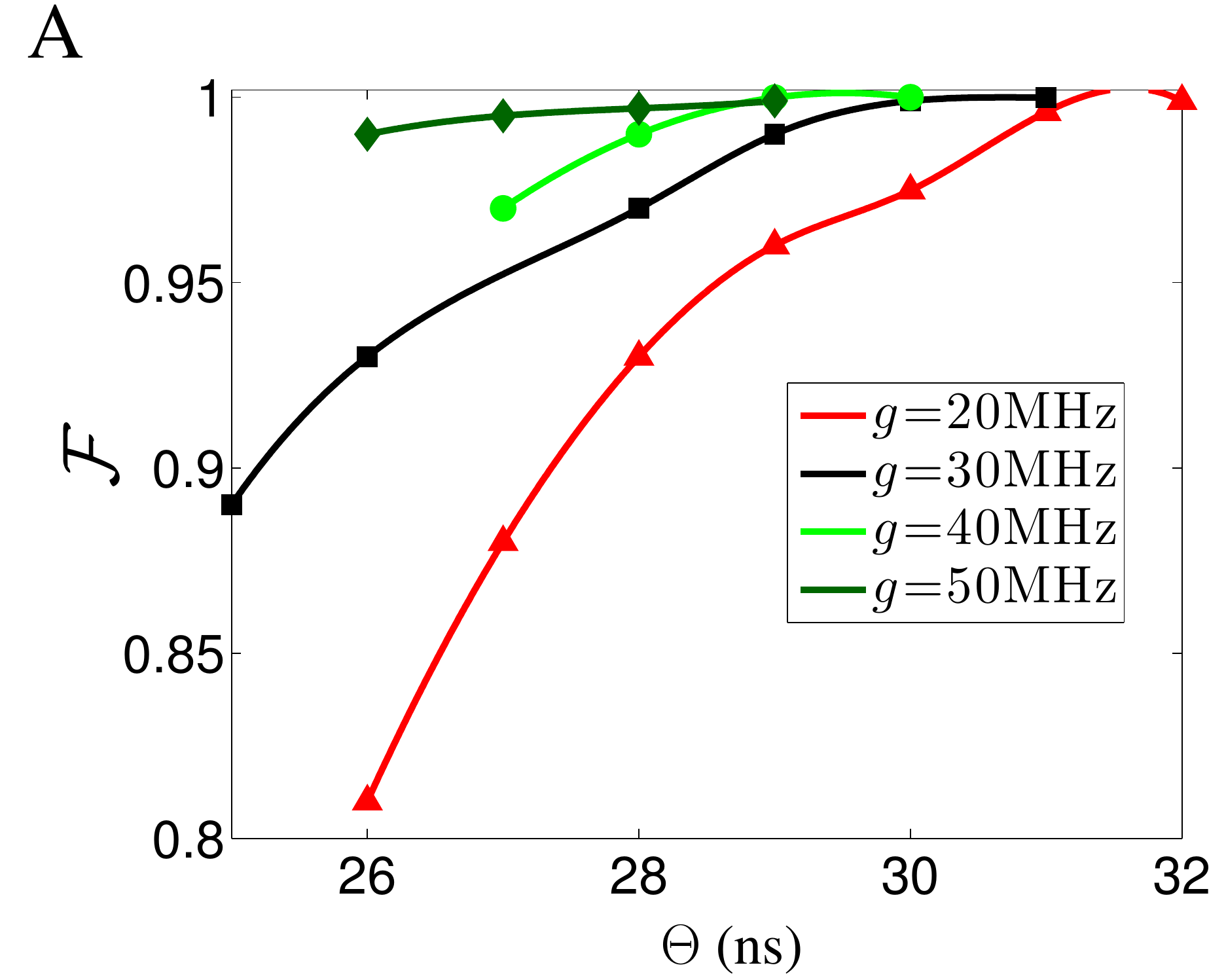}
	\includegraphics[width=0.49\columnwidth]{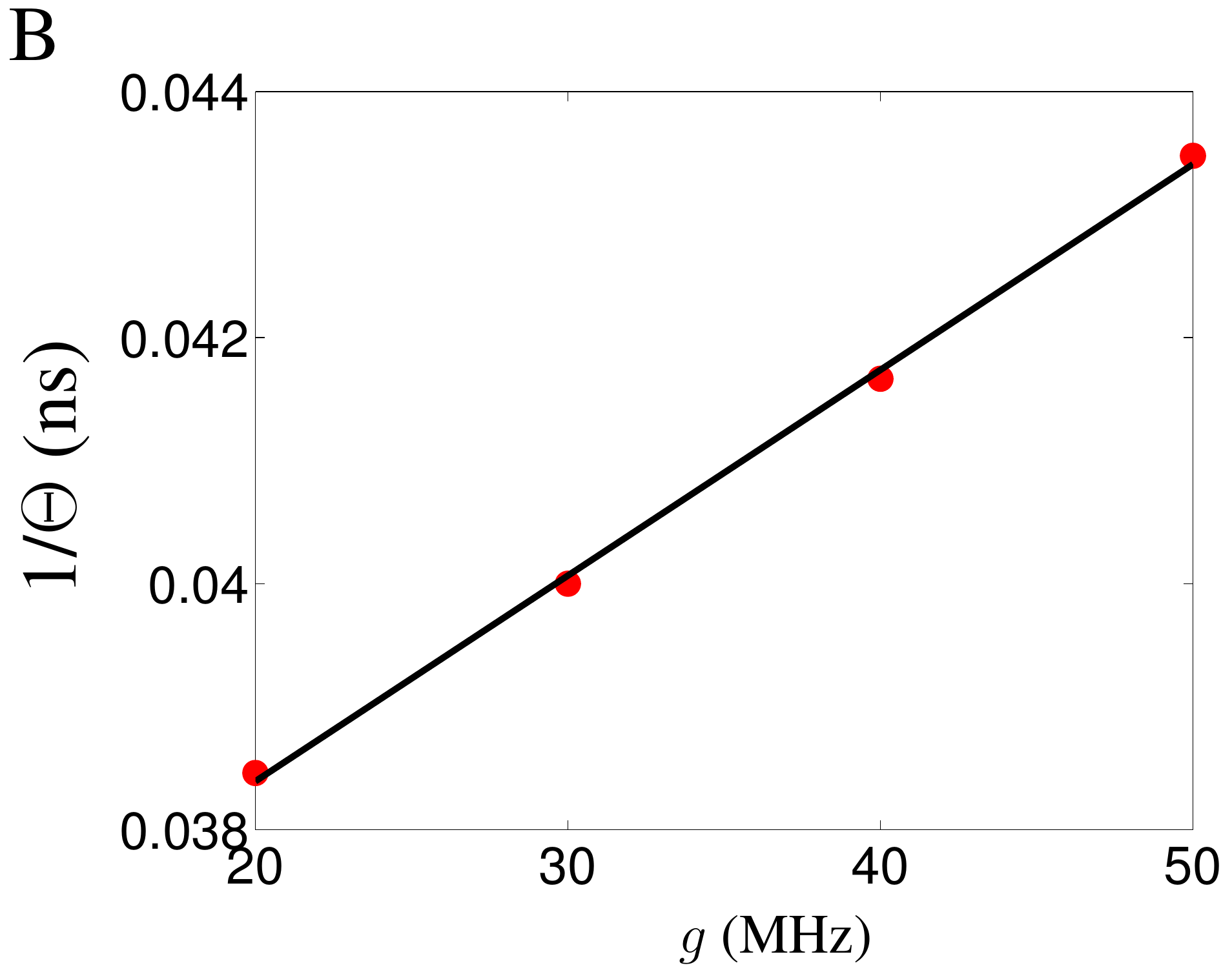}
\caption{
	(color online) A) The dependence of the intrinsic fidelity of the CZZ gate on the evolution time $\tau$ of the system for various values of~$g$. The discretized values show the actual numerical data with $\Diamond$, $\triangle$,  $\circ$, and $\square$ corresponding to
	the values of~$g$ to be 20, 30, 40, 50~\text{MHz}, respectively. A cubic interpolation fits the curves to the data. B) The relation between the inverse of the gate operation time and coupling strength between transmons where the dots
	denote the actual numerical results for various values of $g\in\{20, 30, 40, 50\}$. A linear-fit interpolates the points to the actual data.
	}
\label{fig:C_fidvst}
\end{figure}

We perform the same analysis on the dependence of intrinsic fidelity on the gate operation time, as we did for Toffoli and Fredkin gates.
The analysis is performed for various values of coupling strengths and is shown in Fig.~\ref{fig:C_fidvst}A. The discrete points on the plot show the actual data and the curves are the cubic-fit to the data. Fig.~\ref{fig:C_fidvst}B represents the relation between the inverse of the gate operation time and coupling strength. The discrete points on the plot show the actual numerical data and the solid line is the linear fit to data.

\begin{figure}
	\includegraphics[width=0.9\columnwidth]{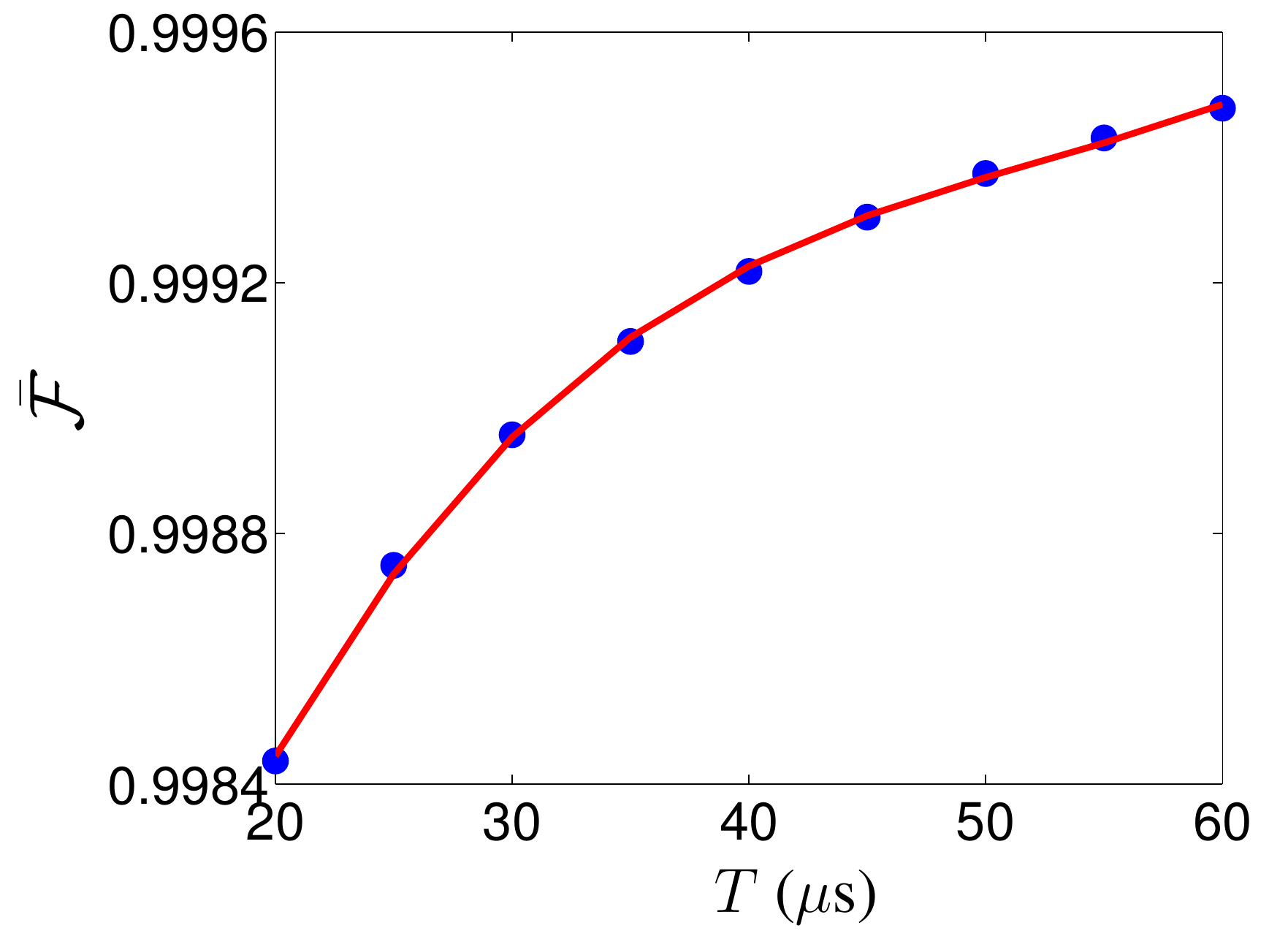}
\caption{
	(color online)
	Fidelity vs coherence time for the CZZ gate. The dots denote the actual numerical data and the blue solid line shows a cubic-fit interpolation on the actual data.}
\label{fig:C_decoherence}
\end{figure}

Figure~\ref{fig:C_decoherence} shows the decoherence induced noise for the CZZ gate.
We plot fidelity vs transmon coherence time.
The actual discrete points are connected via a linear interpolation.
Similar to the Toffoli and Fredkin gates, the decoherence appears in terms of the amplitude- and phase-damping on the transmons. Here the coupling strength is $g=30~\text{MHz}$. 

\begin{figure}
	\includegraphics[width=0.9\columnwidth]{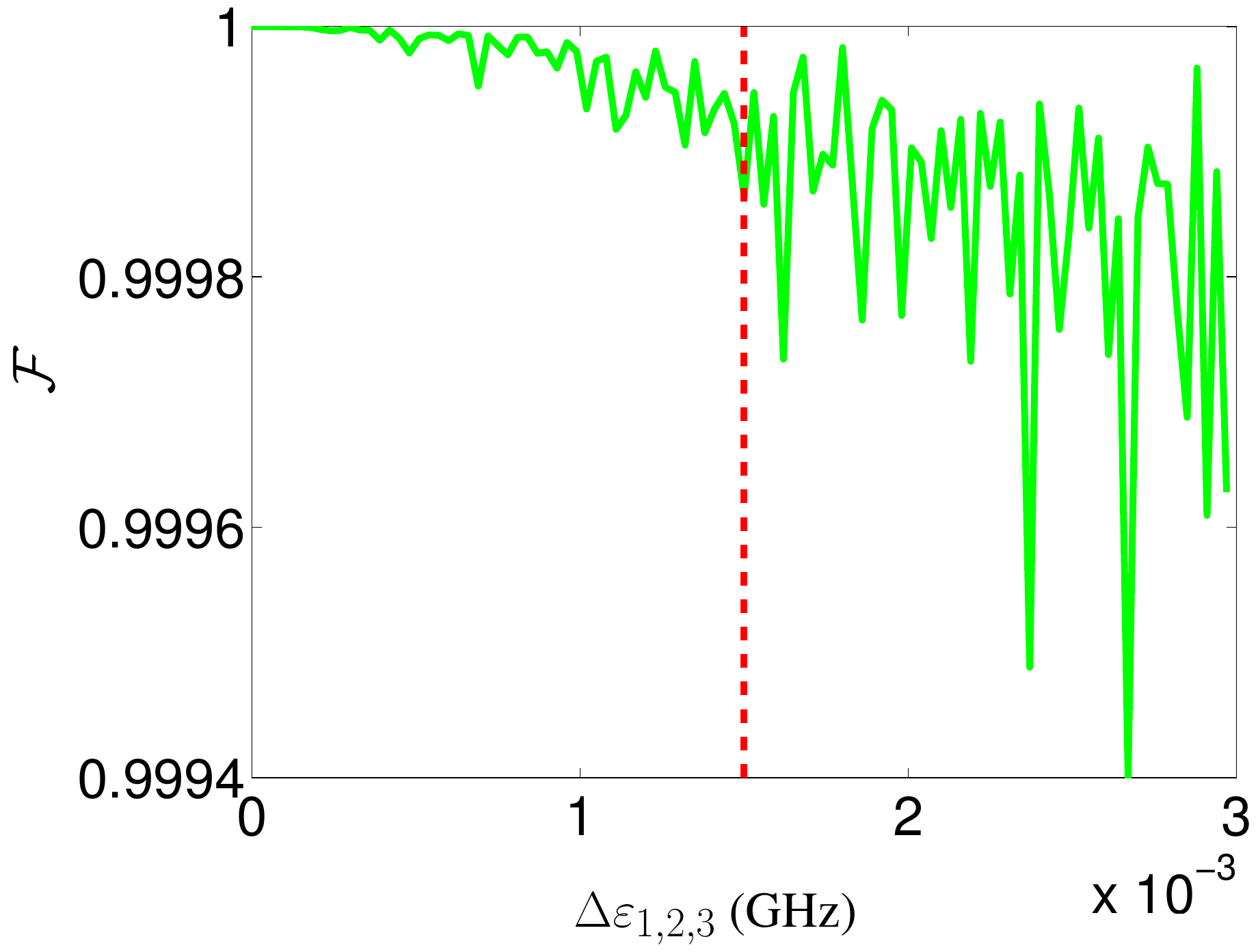}
\caption{Intrinsic fidelity $\mathcal{F}$ vs~$\delta\varepsilon$ for the CZZ gate.
The vertical red dotted-line denotes the threshold, such that on the left side of this line $\mathcal{F}>0.9999$.}
\label{fig:C_robustness}
\end{figure}

We follow the same procedure as for the Toffoli and Fredkin gates to test the robustness of our designed pulse for the CZZ gate. We employ the optimal pulse in Fig.~\ref{fig:C_recp} and apply random noise on each learning parameter.
Figure~\ref{fig:C_robustness} shows the effect of such random noise on the intrinsic fidelity of the CZZ gate which operates in 31 \text{ns}.

\section{Discussion}
\label{sec:discussion}

\begin{table}[htb]
\centering
\begin{tabular}{| c | c | c | c | c |}
\hline
Gates		& $\Theta$(ns)	& $\mathcal{F}$	&$T$($\mu$s)	& $\mathcal{\overline{F}}$ \\
\hline
Toffoli		& 26  		& 0.9999			& 30		& 0.9992 \\
Fredkin		& 26  		& 0.9999			& 30		& 0.9991 \\
CXX			& 31  		& 0.9999			& 30		& 0.9990 \\
\hline
\end{tabular}
\caption{Comparison of fidelities among various three-qubit gates for $g=30$ MHz. $\Theta$, $\mathcal{F}$, $T$ and $\mathcal{\overline{F}}$ are total gate time (in nanoseconds), intrinsic fidelity (defined by Eq.~(\ref{eq:explicitF})), coherence time (in microseconds) and average state fidelity respectively.}
\label{table:gateComp}
\end{table}

We employed a one-dimensional system comprising three coupled superconducting artificial atoms to design single-shot high-fidelity three-qubit gates, such as Toffoli, Fredkin and CXX. Our results include the optimal pulses for each three-qubit gate, analysis of dependence of the intrinsic fidelity on the physical model parameters, and analysis of performance of these gates under decoherence induced noise. Table \ref{table:gateComp} shows a comparison among the numerical simulation of optimal Toffoli, Fredkin and CXX gates for the same coupling strength $g=30$ MHz. The results indicate that the efficacy of SuSSADE is independent of the desired quantum operations. In this section, we discuss the results for all three gates. We first elaborate on our choice of the control pulses. Then we discuss the dependence of fidelities for three-qubit gates on the physical model parameters, discuss about the effects of noise, and finally compare the SuSSADE scheme against alternative approaches.

\subsection{Control pulses}
To formulate the problem of designing the high-fidelity quantum gates into a learning algorithm we represent the qubit frequency in terms of external control functions (hypothesis).
The choice of the control function is ubiquitous and we only choose two pulse profiles, piecewise-constant and error-piecewise-constant, which are relevant for superconducting control electronics. The learning algorithm
 shapes the external pulses to obtain high-fidelity quantum gates. Each of these control pulses has its own advantage and drawback in terms of the computational resource and practical implementation.
 
 The piecewise-constant control function are computationally less expensive, and on average, a single run of the learning procedure using the square pulse takes an order of magnitude less run-time in contrast to the piecewise-error-function pulses.
Piecewise-constant pulses are easy to generate using the current superconducting control electronics. However, the Gaussian filters connecting the control electronics and the physical qubits cause distortion on the square pulses.
Transmons thus receive a distorted pulse. A smooth pulse must, therefore, be generated to account for the first order of distortion numerically.

The piecewise-error-function connects the control parameters smoothly such that the function approximates the realistic control pulses for transmon system. In this way we overcome the problem of infinite bandwidth of the square pulses as well as the first order distortion.
We have numerically shown that (See~\cite{ZGS15} and Figs.~\ref{fig:F_recp} and \ref{fig:C_recp}) the learning procedure does not depend on the shape of the control pulse
but rather depends on the number of learning parameters.

For designing each high-fidelity three-qubit gate, we used the same number of learning parameters either for the piecewise-constant or piecewise-error-function pulses. For Toffoli and Fredkin gate, we used $3\times26$ parameters to design the gates, which operate on 26~\text{ns} using either
piecewise-constant or piecewise-error-function pulses (See~\cite{ZGS15} and Fig.~\ref{fig:F_recp}). The number of learning parameter to design a CZZ gate (Fig.~\ref{fig:C_recp}), which operates over $31$~\text{ns}, is $31~\times3$ for both piecewise-constant and
piecewise-error-function pulses.

\subsection{Intrinsic fidelity}
In~\cite{ZGS15} and Figs.~\ref{fig:F_fidvst}A, and ~\ref{fig:C_fidvst}A, we plot the intrinsic fidelity as a function of the gate operation time for various values of $g\in\{20, 30, 40, 50\}$MHz. Keeping the value of~$g$ fixed, the fidelity is a monotonically increasing function of the gate operation time. This is consistent with the notion that  
 increasing the evolution time of the system increases the fidelity between the unitary evolution and the target gate~\cite{ZGS15}. As the coupling strengths become stronger, the gate operation time to reach
to a threshold fidelity becomes shorter. This is consistent with the idea of avoided-crossing-based gates (\ref{eq:CZtime}). For smaller values of~$g$, the gate operation time becomes longer, and fixing the time-bin
to $1~\text{ns}$, the number of learning parameters increase. Our learning algorithm still delivers the high-fidelity gates despite the increase of the learning parameters.

In~\cite{ZGS15} and Figs.~\ref{fig:F_fidvst}B, and~\ref{fig:C_fidvst}B, we have shown that a linear relation between the inverse of the gate operation time and coupling strength exists.
Providing a theoretical framework to explain
this relation is a challenging task for three-qubit gates. However, from our discussion for avoided-crossing-based CZ gate (Sec.~\ref{subsec:CZ}), one can intuitively expect a linear relation between the coupling strength and the gate operation time for a fixed intrinsic fidelity, even for three-qubit avoided-crossing-based gates. This linear relation also emphasizes the efficacy of the learning algorithm to discover the correct relation between the coupling strength and gate operation time.

\subsection{Noise}
\label{sec:noise}
In this section, we explain the effect of noise on the system evolution of each three-qubit gate in three parts. In the first part, we discuss the effect of random noise on the system and discuss the robustness of our procedure. In the second part, we explain the effects of decoherence induced noise, which are caused by the environment. In the last part of this section, we discuss how the higher orders of noise on the generated procedure can be suppressed.

\subsubsection{Robustness}
We test the robustness of our procedure for designing the three-qubit gates by applying random noise on the learning parameters. Our analysis shows that the devised procedures to design Toffoli, Fredkin and CZZ gates are robust against the external random noise (Figs.~\ref{fig:T_robustness},~\ref{fig:F_robustness},~\ref{fig:C_robustness}), if the magnitude of the random noise is lower than 800, 600 and 1500 KHz respectively, which are within the limit of current state-of-the-art superconducting control-electronics~\cite{CommunicationWithPedram}.

\subsubsection{decoherence induced noise}
The amplitude-damping and dephasing rates ($T_1^{-1}$ and $T_2^{-1}$) determine the decoherence rate of our three-transmon system. Assuming $T=T_1=T_2$ for all the tunable transmon devices, we can plot $\bar{\mathcal{F}}$ versus the coherence time of the transmons (See Ref.\cite{ZGS15} and Figs.~\ref{fig:F_decoherence},~\ref{fig:C_decoherence}). For a gate operation time much faster than the coherence time of the transmons, i.e., $T\gg{\Theta}$,
if the intrinsic fidelity is much smaller than the fidelity,
with this reduction caused by decoherence induced errors,
we can approximate
\begin{equation}
\label{eq:approxi_noise}
	\bar{\mathcal{F}}\sim1-\frac{\Theta}{T}.
\end{equation}
Equation~(\ref{eq:approxi_noise}) matches our numerical simulation of decoherence in~\cite{ZGS15} and Figs.~\ref{fig:F_decoherence}~\ref{fig:C_decoherence}.

When the coherence time of transmon is significantly higher than the gate operation time i.e., $T\gg{\Theta}$, the error from the intrinsic fidelity is the main source of noise. Under this
condition, the intrinsic fidelity must be so high, such that the resultant gate fidelity meets the threshold fidelity for fault-tolerant quantum computing. With the long coherence time (20 $\sim$ 60$\mu\text{s}$) of the state-or-the-art superconducting artificial atoms accompanied with
the machine learning approach that delivers a gate with $\bar{\mathcal{F}}>0.999$, our proposal enables the implementation of high-fidelity three-qubit gates under current experimental conditions.
 
\subsubsection{Distortion of control pulses}
\label{subsec:Cpulsedist}
The learning algorithm can shape any type of external pluses to design high-fidelity three-qubit gates. We employed the piecewise-error-function pulse to resolve the infinite bandwidth problem of the square pulses and to account for
the first-order distortion on the pulse. However, higher order distortions can change the optimal shape of the designed pulse and can lead to a sub-threshold fidelity. For example, we ignored the weak dependence of~$g$ and~$\eta$ on the transmon frequency.

This frequency dependence of physical parameters can introduce small perturbations into the system Hamiltonian, thereby distorting the optimal pulses. One viable option to suppress the higher degree of distortions on the learning parameters is closed-loop learning control~\cite{RVM+00,JR92}, which can be used in conjunction with our control scheme.

\subsection{Comparison against alternative approaches}

\begin{table}[htb]
\centering
\begin{tabular}{| c | c |}
\hline
Method              & $\mathcal{F}_{\text{best}}$ \\
\hline
Quasi-Newton	& 0.9912  \\
Simplex		& 0.9221 \\
DE			& 0.9931 \\
SuSSADE		& 0.9999 \\
\hline
\end{tabular}
\caption{Comparison against alternative approaches of designing Toffoli gates for $g=30$ MHz and $\Theta=26$ ns.}
\label{table:comp}
\end{table}
Table \ref{table:comp} shows a comparison among various approaches of generating the optimal pulse shapes for the three-qubit Toffoli gate. $\mathcal{F}_{\text{best}}$ denotes the best intrinsic fidelity,
defined by Eq.~(\ref{eq:explicitF}),
obtained for the corresponding method.
Note that we have compared SuSSADE against both the greedy optimization algorithms (Quasi-Newton and Simplex) as well as against the global optimization algorithms (DE). The comparison data shown in Table \ref{table:comp} indicates the efficacy of SuSSADE over the alternative approaches.

\section{Conclusion}
\label{sec:conclusion}
In conclusion, we have transformed a problem of designing three-qubit gates into a quantum-control problem. The control problem is then mapped into a supervised machine learning algorithm. In the context of supervised machine learning, we used the truth-table data for each quantum gate as the training data. The transmon frequencies represent the hypothesis. The supervised learning algorithm then trains the qubit frequencies on the truth table data to generate procedures for designing high-fidelity three-qubit gates. Our approach to defining a gate-design problem as a supervised machine learning can inspire the application of other supervised machine learning methods, such as Support Vector Machine~\cite{Hay98} and Neural Network~\cite{CV95} for designing quantum gates which acts on more than three qubits.

We have already introduced the quantum control scheme, named Subspace-Selective Self-Adaptive Differential Evolution (SuSSADE), in designing high-fidelity Toffoli gate~\cite{ZGS15}. Here, we employ the SuSSADE algorithm to generate procedures for other three-qubit gates, such as Fredkin and CXX gates.
The two three-qubit gates considered here operate as fast as the two-qubit entangling CZ gate under the same experimental constraints. The robust performance of these gates against decoherence induced as well as random errors signifies the efficacy and robustness of the SuSSADE scheme for optimizing promising superconducting architectures.
The system considered here comprises three nearest-neighbor-coupled transmons that can serve as a module for any 1D or 2D architecture, and, in fact, our three-qubit gates can be realized in such multi-qubit systems, if the undesired couplings are switched off.
Our results here establish the efficacy of SuSSADE as a machine learning approach to designing high-fidelity quantum operations.

\begin{acknowledgments}
This research was partially funded by NSERC and Alberta Innovates.
EZ appreciates financial support from MITACS.
JG acknowledges financial support from the University of Calgary's Eyes High Fellowship Program
and was also supported by ONR under award no.\  N00014-15-1-0029
and by NSF under award no.\ PHY-1104660.
BCS appreciates financial support provided by China's 1000 Talent Plan
and by the Institute for Quantum Information and Matter, 
which is a National Science Foundation Physics Frontiers Center
(NSF Grant PHY-1125565)
with support of the Gordon and Betty Moore Foundation
(GBMF-2644).
This research was enabled in part by support provided by WestGrid (www.westgrid.ca) and Compute Canada Calcul Canada (www.computecanada.ca).
We thank Alexandre Blais, Austin Fowler, Michael Geller and Simon Nigg for valuable discussions
and Jonathan Johannes for proofreading the manuscript.
\end{acknowledgments}
%
%%%%%%%%%%%%%%%%%%%%%%%%%%%%%%%%%%%%%%%%%%%%%%%%%%%%%%%%%%%%%%%%%%%%%%%%%%%%%%%%%%%%
%
%\bibliographystyle{apsrev}
\bibliography{3qubit_gates}

\end{document}